\documentclass[published]{agujournal2025}

\usepackage{booktabs}
\usepackage{amsmath}
\usepackage{multirow}
\usepackage{url}

\begin{document}

\journalname{Journal of Geophysical Research: Space Physics}

\title{Cosmic Ray Inter-Station Fluctuations as Precursors of Geomagnetic Storms: Causal Baseline Analysis and Forecasting Limitations}

\authors{
 Haoyang Li$^{*}$\affil{1},
 Zongyuan Ge$^{*}$\affil{2},
 Zhaoming Wang$^{\dagger}$\affil{1,3,4}
}

\affiliation{1}{College of Physics and Optoelectronic Engineering, Ocean University of China, Qingdao 266100, China\newline}
\affiliation{2}{Department of Physics, National University of Singapore, Singapore 117551, Singapore\newline}
\affiliation{3}{Engineering Research Center of Advanced Marine Physical Instruments and Equipment of Ministry of Education, Qingdao 266100, China\newline}
\affiliation{4}{Qingdao Key Laboratory of Optics and Optoelectronics, Qingdao 266100, China}

\authoraddr{
 Zhaoming Wang, Ocean University of China, Qingdao 266100, China (wangzhaoming@ouc.edu.cn).
}

\authorrunninghead{Li, Ge, and Wang}
\titlerunninghead{Cosmic Ray Precursors of Geomagnetic Storms}

\renewcommand{\firstauthorinitialedname}{LI, GE, AND WANG}
\renewcommand{\articlenumber}{10.1029/2025JAXXXXXX}

\keypoints
 {A strictly causal baseline and direction-free fluctuation indices are used to analyze 25 years of neutron monitor data.}
 {Both indices are enhanced, on average, in the days before storm onset, significantly above a quiet random control.}
 {A single-threshold alert does not exceed the random baseline; the indices are a descriptive precursor, not an intensity forecaster.}

\renewcommand{\acknowledgementstext}{
We thank Fanghua Liu for helpful discussions. This work was supported by the Natural Science Foundation of Shandong Province (Grant Nos. ZR2024MA046 and ZR2021LLZ004) and the Fundamental Research Funds for the Central Universities (Grant No. 202364008). We thank the Neutron Monitor Database (NMDB; EU FP7 Grant No. 213007) and the Principal Investigators of the neutron monitor stations Tsumeb (Max Planck Institute for Aeronomy), Potchefstroom (North-West University, South Africa), Jungfraujoch (University of Bern, Switzerland), Kerguelen (Observatoire de Paris, France), Oulu (Sodankylä Geophysical Observatory, Finland), Thule (Bartol Research Institute, University of Delaware, USA), and South Pole (University of Wisconsin--River Falls, USA) for providing and maintaining the data used in this study.
}

\maketitle

\vspace{6pt}
\noindent\rule{\textwidth}{0.4pt}\par
\vspace{4pt}
\small
* These two authors contributed equally to this work.\par
\vspace{2pt}
$\dagger$ Contact author: wangzhaoming@ouc.edu.cn
\vspace{10pt}
\normalsize

\begin{abstract}

The modulation of galactic cosmic rays (GCRs) by interplanetary disturbances, manifested as Forbush decreases, provides a potential signature of coronal mass ejection passage. Previous precursor studies often suffer from solar-cycle bias due to non-causal baselines and lack strict false-alarm evaluations. Here we analyze 25 years of hourly neutron-monitor data from seven stations (1995--2019) using a strictly causal baseline fitted only to past geomagnetically quiet days. We introduce two direction-free metrics: the inter-station dispersion ($A_{\mathrm{disp}}$) and the mean-removed variance ($A'_{\mathrm{var}}$). Using superposed-epoch analysis aligned on storm sudden commencements (SSCs), we find that both indices are enhanced, on average, in the days before storm onset relative to a quiet random control, and that the window-mean amplitude increases with storm intensity. The enhancement is broad and should not be interpreted as an event-specific lead time. A Molchan-diagram evaluation shows that a single-threshold alert based on $A_{\mathrm{disp}}$ or $A'_{\mathrm{var}}$ does not exceed the random-alert baseline at the 95th-percentile threshold. The correlation with storm intensity is moderate ($\rho=-0.28$; AUC $=0.66$ for severe versus moderate storms). These results indicate that inter-station cosmic-ray fluctuations are useful diagnostics of the disturbed heliospheric state, but they are not an independent operational forecasting tool; a reliable warning system will require combination with solar-wind and interplanetary-magnetic-field information.

\begin{plainlanguagesummary}

Disturbances from the Sun can trigger magnetic storms around Earth that affect satellites and power grids. High-energy particles called cosmic rays can sense these disturbances before they arrive, offering a possible early signature. In this study, we use 25 years of cosmic-ray data from seven ground stations to see how the counts change before storms. To make sure the method can be used in real time, we use only past data to define the quiet baseline. We find that differences in cosmic-ray counts between stations are higher, on average, in the days before a storm. However, the increase is broad and gradual, not a sharp countdown, and a simple automated warning system based on this signal produces many false alarms and cannot predict how strong a storm will be. Our results suggest that cosmic rays can tell us that the heliosphere is disturbed, but they must be combined with direct solar-wind measurements to build a reliable forecasting system.
\end{plainlanguagesummary}
\end{abstract}

\section{Introduction} \label{sec:intro}

Geomagnetic storms (GSs) represent one of the most significant manifestations of space weather, arising from the interaction between solar wind disturbances and Earth's magnetosphere \cite{Gonzalez1994}. These events pose substantial risks to technological infrastructure, including satellite operations, power grid stability \cite{Pedersen2024}, and navigation systems \cite{Schrijver_Siscoe_2010}. GSs are driven by two primary solar wind structures: coronal mass ejections (CMEs) \cite{Richardson2012} and high-speed streams (HSSs) emanating from coronal holes \cite{Tsurutani1995}. CMEs are transient, explosive eruptions that carry massive magnetized plasma structures through interplanetary space \cite{Schwenn2006}, typically reaching Earth within 1--5~days depending on their initial speed and propagation conditions \cite{Gopalswamy2001}. In contrast, HSSs are quasi-stationary structures that corotate with the Sun, exhibiting characteristic 27-day recurrence patterns; they interact with the preceding slow solar wind to form corotating interaction regions (CIRs) that can also drive geomagnetic activity \cite{Tsurutani1995,Richardson2018}.\\
The two driver types produce distinct GS characteristics. CME-driven storms tend to be more intense and are responsible for nearly all major storms (Dst $\leq -100$~nT), while CIR/HSS-driven storms are typically moderate (rarely exceeding Dst $\sim -100$~nT) but more frequent during solar minimum \cite{Richardson2012,Zhang2007}. Crucially, the galactic cosmic ray (GCR) response also differs: CME-associated Forbush decreases (FDs) are generally deeper and more abrupt due to the strong magnetic barriers of CME flux ropes and driven shocks. Specifically, when a CME's magnetic field contains a strong southward component, magnetic reconnection at the dayside magnetopause efficiently transfers solar wind energy into the magnetosphere, intensifying the ring current and producing characteristic Dst index depressions \cite{Gonzalez1994}, while CIR-associated decreases are shallower and more gradual \cite{Tsurutani1995,Cane2000,Dumbovic2011}. Our sample includes both CME- and CIR/HSS-driven storms. Because CMEs dominate the severe and extreme classes, the composite signal is expected to be conservative for CME-driven events; we do not separate the two drivers in this work, and we return to this caveat in Section~\ref{sec:discussion}.\\
Current operational space weather forecasting relies on a well-established causal chain: solar eruption observation, interplanetary propagation modeling, and magnetospheric response prediction \cite{tsurutani1997interplanetary}. Coronagraph observations of CMEs can identify potential threats several days in advance, while in-situ solar wind measurements at the first Lagrangian point (L1)---primarily from the Advanced Composition Explorer (ACE) and Deep Space Climate Observatory (DSCOVR)---provide 15--60 minutes of warning before magnetospheric impact \cite{Stone1998,Lotoaniu2022}. Empirical and physics-based models relating solar wind parameters to geomagnetic indices have achieved considerable success \cite{Temerin2002,Bala2012}. Recent machine-learning studies have improved short-term Dst prediction and reduced forecast lag, but these models remain constrained by near-Earth solar-wind and IMF inputs and by the L1-to-magnetopause transit bound \cite{Zhou2026JGR,Zhou2026RAA}. Physics-informed models that combine GCR modulation with solar-wind inputs have begun to improve storm forecasting \cite{Ge2026}. Nevertheless, CME propagation uncertainties produce false alarms or missed events, and extreme events remain difficult to predict because of sample scarcity \cite{Gonzalez1994}.
Galactic cosmic rays (GCRs) offer a complementary pathway to extend geomagnetic storm (GS) prediction windows. As high-energy charged particles ($\sim$GeV) traversing the heliosphere, GCRs have gyroradii comparable to or smaller than the scale of CME-related magnetic structures, making them susceptible to modulation by large-scale interplanetary magnetic field (IMF) structures, and they can respond rapidly to field variations\cite{Kojima2025}. The enhanced magnetic field and turbulent structures associated with CME-driven shocks and magnetic clouds scatter and deflect GCRs, producing measurable intensity variations, known as Forbush decreases (FDs), before and during geomagnetic storms \cite{Cane2000}. This phenomenon was first systematically documented by \citet{Forbush1938} and is often accompanied by energetic particle signatures tied to shock acceleration \cite{Bell1978,Belov2009}. Using wavelet transform and cross-correlation analysis of neutron monitor data from KIEL and MOSC stations during 2003--2004, \citet{Idosa2023} reported robust negative correlations between cosmic ray flux and Dst.\\
Beyond isotropic flux changes, GCR anisotropy provides additional diagnostic information. Classical transport theory predicts anisotropic streaming driven by spatial gradients and magnetic field topology \cite{Forman1975,Schlickeiser2019}, and early observations reported transient north--south anisotropy during interplanetary disturbances \cite{Duggal1976}. \citet{Bieber1998} demonstrated that the anisotropy enhancement preceding CME magnetic cloud arrivals originates primarily from $B \times \nabla n$ drift, where density gradients perpendicular to the IMF drive directional particle flows. This mechanism has been confirmed by subsequent studies \cite{Munakata2005}. \citet{Belov1995} identified precursor signals appearing hours to a day before FD onset, attributable to shock acceleration and loss-cone effects. More recently, \citet{Zhu2015} developed wavelet-based methods detecting 12-hour period enhancements before large storms, while \citet{Grigoryev2017} reported prediction probabilities using band harmonic analysis of GCR angular distributions. CME-type dependent GCR modulation has also been reported in statistical studies \cite{Mishra2008}.\\
Recent investigations have further quantified the coupling between solar wind parameters and cosmic ray intensity (CRI) modulation. For instance, \citet{Uga2023CRI} analyzed data from different neutron monitor stations and combined it with solar wind parameters and geomagnetic indices, confirming in a study spanning multiple observational cycles that variations in CRI not only respond to global solar-terrestrial disturbances but also exhibit significant spatial variation characteristics across stations. \citet{Uga2025} employed Superposed Epoch Analysis and Wavelet Transform Coherence to demonstrate a strong inverse correlation between CRI and key coupling parameters during intense storms, with the coherence being most pronounced during super-storm events. This underscores the importance of storm intensity and solar wind drivers in shaping GCR responses, yet an integrated early-warning framework utilizing multiple GCR characteristics remains undeveloped.\\
Despite these advances, several gaps remain. First, many key results derive from case studies rather than systematic large-sample statistical analyses, making it difficult to quantify precursor signal universality and stability \cite{Belov1995,Munakata2005}. Second, anisotropy quantification methods vary widely, from complex spherical harmonic inversions to qualitative multi-station comparisons, hindering operational implementation \cite{Bieber1998}. Third, integrated frameworks combining multiple GCR characteristics for graded warnings across different storm intensities have not been developed \cite{Grigoryev2017,Zhu2015}.\\
\\
In this paper, we address these limitations by adopting a strictly causal baseline and focusing on direction-free inter-station fluctuation indices. We compute the baseline for each station using only past geomagnetically quiet days, ensuring the analysis can be reproduced in real time. To quantify the spatial non-uniformity of GCR modulation without relying on directional assumptions, we introduce two simplified metrics: the inter-station dispersion (\(A_{\mathrm{disp}}\)) and the mean-removed variance (\(A'_{\mathrm{var}}\)). Using 25 years of data from seven neutron monitor stations (1995--2019), we systematically evaluate the ensemble-mean precursor behavior, the statistical significance against random epochs, and the PCA-based diagnostic consistency of these metrics. Crucially, we also perform a rigorous assessment of their event-level forecasting skill and intensity correlation. Our results demonstrate that while these direction-free indices provide a robust diagnostic of the disturbed heliospheric state preceding storms, they do not achieve operational event-level warning skill.\\

\section{Data and Methods} \label{sec:data}

\subsection{Observational Data} \label{subsec:data}

We utilize hourly pressure-corrected count rates from seven NM stations in the Neutron Monitor Database (NMDB; \url{http://www.nmdb.eu}): OULU (Finland, 65.1$^{\circ}$N), JUNG (Jungfraujoch, Switzerland, 46.5$^{\circ}$N), SOPO (South Pole, 90.0$^{\circ}$S), THUL (Thule, Greenland, 76.5$^{\circ}$N), KERG (Kerguelen, 49.4$^{\circ}$S), PTFM (Potchefstroom, South Africa, 26.7$^{\circ}$S), and TSMB (Tsumeb, Namibia, 19.2$^{\circ}$S). NM response characteristics and rigidity dependence are described in \citet{Clem2000}. The study period extends from 1995 January to 2019 December, encompassing solar cycles 23 and 24.\\
Station coverage is not uniform. OULU and JUNG are more than 96\% complete; THUL and KERG exceed 94\%; SOPO (79.8\%), PTFM (74.7\%), and TSMB (83.3\%) have multi-year or year-scale gaps. Specifically, SOPO has no data in 2006--2009, PTFM in 2012--2015 (and only 8.8\% in 2011), and TSMB has reduced coverage in 2005 (25.6\%), 2017 (62.9\%), and 2019 (18.3\%); KERG has 73.3\% in 2016. The 25-year record contains 219,144 hourly slots. Hours with valid data from at least four stations are retained; 96.1\% of hours have at least five stations and 48.2\% have all seven. Hours with fewer than four stations (3.9\%) are excluded from the network analysis.\\

\begin{table}[htbp]
\centering
\small
\caption{Neutron-monitor stations used in this study. Coordinates, altitudes, and effective vertical cutoff rigidities are from the NMDB station table. Coverage is the fraction of hourly slots in 1995--2019 with valid data after quality control.}
\label{tab:stations}
\begin{tabular}{lcccccc}
\toprule
Station & Latitude & Longitude & Altitude (m) & Cutoff (GV) & Detector & Coverage (\%) \\
\midrule
OULU & $65.05^{\circ}$N & $25.47^{\circ}$E & 15 & 0.8 & 9-NM64 & 96.1 \\
JUNG & $46.55^{\circ}$N & $7.98^{\circ}$E & 3570 & 4.5 & 18-IGY & 96.1 \\
SOPO & $90.00^{\circ}$S & --- & 2820 & 0.1 & $3\times{}^{3}$He NM64 & 79.8 \\
THUL & $76.50^{\circ}$N & $68.70^{\circ}$W & 26 & 0.3 & 9-NM64 + BF3 & 95.9 \\
KERG & $49.35^{\circ}$S & $70.25^{\circ}$E & 33 & 1.14 & 18-NM64 & 94.7 \\
PTFM & $26.68^{\circ}$S & $27.09^{\circ}$E & 1351 & 6.94 & 15-IGY & 74.7 \\
TSMB & $19.20^{\circ}$S & $17.58^{\circ}$E & 1240 & 9.15 & 18-NM64 & 83.3 \\
\bottomrule
\end{tabular}
\begin{flushleft}
\footnotesize Note. The cutoff rigidity is highest at TSMB (9.15~GV) and second highest at PTFM (6.94~GV). Coverage gaps are discussed in Section~\ref{subsec:limits}.
\end{flushleft}
\end{table}

Geomagnetic activity is characterized using the hourly Dst index from the World Data Center for Geomagnetism, Kyoto. Storms are classified into four intensity levels based on minimum Dst: minor ($-50 < \mathrm{Dst_{min}} \leq -30$~nT; 496 events), moderate ($-100 < \mathrm{Dst_{min}} \leq -50$~nT; 378 events), severe ($-200 < \mathrm{Dst_{min}} \leq -100$~nT; 93 events), and extreme ($\mathrm{Dst_{min}} \leq -200$~nT; 17 events). A total of 286 storms have a cataloged SSC, and 208 of these have Dst$_{\min} \leq -50$~nT.\\
Data preprocessing includes: (1) isolated spikes are removed if they deviate from the 49-hour running median by more than 8\% or fall outside the range of 0.5 to 1.5 times the global median; (2) linear interpolation is applied to isolated missing values with a maximum gap length of 24 consecutive hours, while gaps exceeding 24 hours are left as missing and excluded from further analysis. Hours with valid data from fewer than four stations are excluded from the network analysis; and (3) temporal alignment across all data sources.\\

\subsection{Analysis Framework} \label{subsec:framework_method}

Our analysis proceeds in three stages, addressing the two key questions: (1) How do GCR characteristics correlate with GS intensity? and (2) Can these correlations provide early warning signals?\\
\\
\textbf{Stage 1: Ensemble-mean precursor characterization.} Superposed-epoch analysis (SEA) aligned on the storm sudden commencement (SSC) is adopted to derive ensemble-mean temporal profiles of cosmic-ray metrics for four storm-intensity classes. The SSC is used instead of the Dst minimum because the main-phase Forbush decrease would otherwise contaminate the precursor window. Bootstrap resampling is applied to compute 95\% confidence intervals. We further perform a random-epoch Monte Carlo analysis to establish the statistical significance: for each event, a random epoch is drawn from the same calendar year, avoiding any storm or SSC within $\pm 5$~days, and the superposed analysis is repeated 500 times to obtain the 95\% envelope. The precursor is considered significant when the observed curve continuously exceeds the upper envelope.\\
\\
\textbf{Stage 2: Fluctuation assessment.} We evaluate whether inter-station fluctuation features provide precursor signals beyond single-station flux changes. Two complementary approaches are employed: a data-driven method based on principal-component analysis (PCA) of multi-station residuals, and a physically motivated method based on simplified inter-station fluctuation indices (\(A_{\mathrm{disp}}\) and \(A'_{\mathrm{var}}\)) introduced here for operational monitoring.\\
\textbf{Stage 3: Case validation.} Representative extreme and severe storm events are analyzed to verify that the statistically identified precursor patterns manifest in individual cases.\\

\subsection{GCR Characteristic Parameters} \label{subsec:parameters}

The raw neutron monitor count rates have been corrected for atmospheric pressure. To remove the long-term trends caused by the solar activity cycle and interannual variations, we construct a strictly causal baseline for each station. For each day, the baseline is fitted from the daily means of the preceding 365 days that are geomagnetically quiet (all-day $|\mathrm{Dst}| \le 30$~nT), using tricube-weighted local linear regression and extrapolated to the current day. The diurnal variation is removed by fitting 24-hour and 12-hour harmonics to the preceding 27 days and subtracting the fitted wave. No future data enter the baseline at any step, so the detrended series can be reproduced in real time.\\For station \(i\), let the hourly count rate be \(C_i(t)\) (where \(t\) is the time index) and \(T_i(t)\) the fitted baseline. The relative deviation at each time is then calculated as:\\
\begin{equation}
R_i(t) = \frac{C_i(t) - T_i(t)}{T_i(t)} \times 100\% \qquad
\end{equation}
\(R_i(t)\) represents the percentage deviation of the current count rate from the quiet background under the same solar activity phase, with negative values corresponding to decreases and positive values indicating relative enhancements. This normalization removes systematic differences among stations, isolating differential modulation effects.\\
We define two three parameters to characterize GCR variations:
\textbf{Network variance ($A_{\mathrm{var}}$).} We introduce a simplified fluctuation index for operational monitoring:\\
\begin{equation}
A_{\mathrm{var}}(t) =\frac{1}{n}\sum_{i=1}^{n} R_i(t)^2
\end{equation}
This quantity is the mean of the squared relative deviations across the network, measuring the total fluctuation amplitude. It increases when the seven stations deviate collectively from their quiet background.\\
\textbf{Mean-removed variance (\(A'_{\mathrm{var}}\)).} The raw network variance \(A_{\mathrm{var}}\) can be decomposed as:\\
\begin{equation}
A_{\mathrm{var}}(t) = A'_{\mathrm{var}}(t) + \bar{R}(t)^2, \quad \text{where} \quad \bar{R}(t) = \frac{1}{n}\sum_{i=1}^{n} R_i(t)
\end{equation}
Here, $\bar{R}^2$ is the network-mean term that reflects the global (isotropic) Forbush decrease, while $A'_{\mathrm{var}}$ is the inter-station component:
\begin{equation}
A'_{\mathrm{var}}(t) = \frac{1}{n}\sum_{i=1}^{n} \left[ R_i(t) - \bar{R}(t) \right]^2
\end{equation}

\textbf{Inter-station dispersion ($A_{\mathrm{disp}}$).} We further define:\\
\begin{equation}
A_{\mathrm{disp}}(t) = \max R_i(t) - \min R_i(t)
\end{equation}
This quantity is the spread between the most and least deviated stations, measuring the degree of inter-station disagreement. Unlike $A_{\mathrm{var}}$, which aggregates contributions from all stations, $A_{\mathrm{disp}}$ captures the two-endpoint spread of the station distribution.\\
We emphasize that both $A_{\mathrm{var}}$ and $A_{\mathrm{disp}}$ are computed from LOWESS-detrended relative deviations. Neither index contains directional information, and neither is interpreted as anisotropy. They are therefore direction-free fluctuation metrics suitable for rapid computation from raw neutron-monitor data.\\
As an unsupervised, data-driven diagnostic of the multi-station residuals $R_i(t)$ \cite{Ugwoke2023}, we perform PCA on the standardized residuals. No prior physical interpretation is assigned to individual components. We use the PCA only to identify the dominant modes and to test whether the residual variance is dominated by a network-wide mode; it is not independent evidence for an inter-station precursor. We compare the PC1 score with the network mean $\bar{R}$ and the PC2 score with $A_{\mathrm{disp}}$, and summarize the differences with Cliff's $\delta$. Significance is assessed with a 96-h block permutation (5000 shuffles), which preserves within-block autocorrelation; an hourly permutation would overstate significance.\\

\subsection{Event-Level Alert Evaluation}\label{subsec:alertdef}
For each index $X(t)$, an alert is issued at hour $t$ if $X(t)$ exceeds a causal threshold $\theta_p(t)$ and the current Dst is above $-30$~nT (i.e., the system is not already in a storm):
\begin{equation}
\theta_p(t)=\mathrm{percentile}_p\{X(s):t-8760\,\mathrm{h}\leq s<t\},\qquad
A(t)=1\ \mathrm{if}\ X(t)>\theta_p(t)\ \mathrm{and}\ \mathrm{Dst}(t)>-30\,\mathrm{nT}.
\end{equation}
The threshold is the $p$-th percentile of the preceding 365 days, updated daily, and requires at least 60 days of prior data. Consecutive alert hours are merged into episodes when the gap between them is shorter than 12~h. The storm-onset target $t_0$ is the main-phase onset (Dst maximum in the 48~h preceding the Dst minimum). An episode is a hit if at least one alert hour falls in $[t_0-48\,\mathrm{h},t_0)$. The probability of detection and false-alarm ratio are
\begin{equation}
\mathrm{POD}=\frac{N_{\mathrm{hit}}}{N_{\mathrm{events}}},\qquad
\mathrm{FAR}=\frac{N_{\mathrm{false}}}{N_{\mathrm{episodes}}},
\end{equation}
where an episode $[a,b]$ is counted as a false alarm if no storm onset occurs in $[a,b+48\,\mathrm{h}]$. The 48-h window is a fixed evaluation horizon; it is not a claim of 48-h predictive skill. The time coverage is the fraction of valid hours for which an alert occurred in the preceding 48~h:
\begin{equation}
C=\frac{1}{N_{\mathrm{valid}}}\sum_t \mathbf{1}\{\mathrm{alert\ in}\ [t-48\,\mathrm{h},t)\}.
\end{equation}
The Molchan diagram plots the miss rate $(1-\mathrm{POD})$ against $C$. A system with no skill lies on the diagonal $1-C$; a skillful system lies below it. The Molchan skill score is $S=0.5-\int(1-\mathrm{POD})\,dC$, so $S>0$ indicates positive skill. We scan thresholds from the 80th to the 99th percentile and report the 95th percentile as the representative operating point. All thresholds are computed from past data only, so the evaluation is causal.

\section{Results} \label{sec:results}

Our results are organized around two questions: (1) Are the inter-station indices enhanced, on average, before storms, and does the enhancement scale with intensity? (2) Does a single-threshold alert based on these indices provide event-level skill? We first characterize the ensemble-mean enhancement and its significance, then test the event-level alert and the intensity correlation.\\

\subsection{Ensemble-Mean Precursor Characterization}
\label{subsec:precursor}

\begin{figure}[htbp]
\centering
\includegraphics[width=0.6\textwidth]{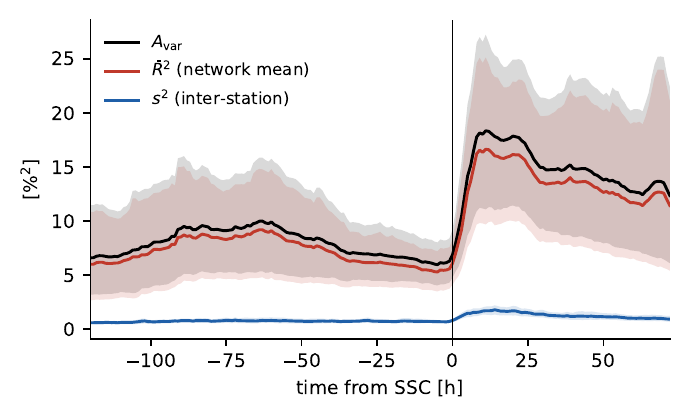}
\caption{Decomposition of the network variance \(A_{\mathrm{var}} = \bar{R}^2 + s^2\) around the shock arrival (SSC). The black curve is the total variance, the red curve is the network-mean term \(\bar{R}^2\), and the blue curve is the inter-station variance \(s^2\). The network-mean term dominates the total variance, especially during the main phase (after SSC). This confirms that the raw \(A_{\mathrm{var}}\) primarily reflects the global Forbush decrease rather than cross-station fluctuation. The mean-removed version \(A'_{\mathrm{var}}\) (or equivalently, \(s^2\)) isolates the spatial heterogeneity and is therefore used as the primary variance index in our analysis.}
\label{fig:avar_decomp}
\end{figure}

To demonstrate why the mean-removed variance is necessary, Figure~\ref{fig:avar_decomp} shows the temporal decomposition of \(A_{\mathrm{var}}\) into the network-mean term \(\bar{R}^2\) and the inter-station variance \(s^2\) aligned on SSC. The network-mean term clearly dominates the total variance, particularly during the storm main phase, which reflects the global Forbush decrease. The inter-station variance \(s^2\) accounts for only a small fraction of the total. Therefore, using the raw \(A_{\mathrm{var}}\) would conflate the spatial non-uniformity with the global cosmic-ray depression. The mean-removed version \(A'_{\mathrm{var}}\) explicitly isolates the inter-station component and is used in the remainder of the analysis.\\

\subsubsection{Intensity-Dependent Precursor Amplitude}\label{subsubsec:Precursor}

\begin{figure}[htbp]
\centering
\begin{minipage}{0.48\textwidth}
 \centering
 \includegraphics[width=\linewidth]{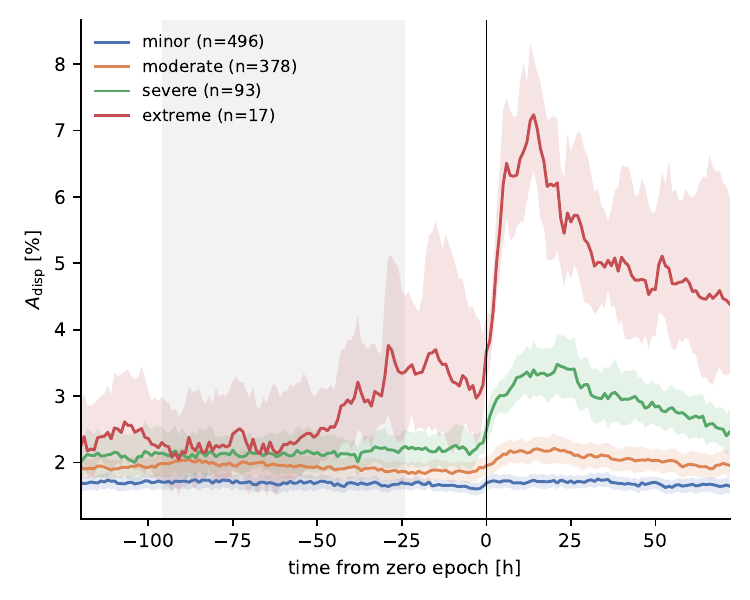}
 \textit{(a) \(A_{\mathrm{disp}}\)}
\end{minipage}\hfill
\begin{minipage}{0.48\textwidth}
 \centering
 \includegraphics[width=\linewidth]{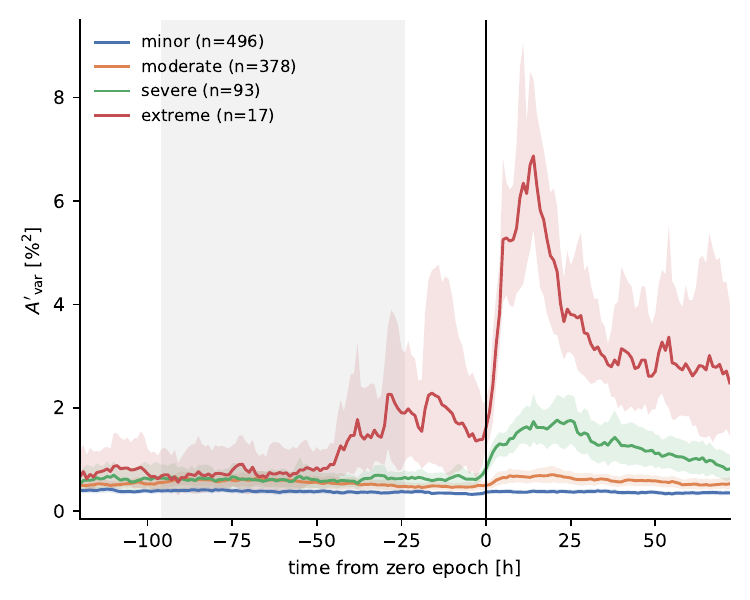}
 \textit{(b) \(A'_{\mathrm{var}}\)}
\end{minipage}
\caption{Ensemble-mean temporal profiles of (a) \(A_{\mathrm{disp}}\) and (b) \(A'_{\mathrm{var}}\) aligned on SSC or main-phase onset, shown separately for four storm-intensity classes. Solid lines indicate ensemble means; shaded regions indicate 95\% confidence intervals from year-block bootstrap. The gray region marks the precursor window ($-96$ to $-24$~h). Both indices show systematic enhancement that scales monotonically with storm intensity. The Extreme class contains only 17 events, which explains the wider confidence interval.}
\label{fig:sea_intensity}
\end{figure}

We first examine the ensemble-mean temporal profiles of the fluctuation indices for the four storm-intensity classes. The zero epoch is the SSC (or the main-phase onset for events without a cataloged SSC), and the Dst minimum is used only as an outcome variable. Aligning on the Dst minimum would contaminate the precursor window with the main-phase Forbush decrease, since 26\% of SSC events fall inside the nominal ``minimum minus 96--24~h'' window.\\Figure~\ref{fig:sea_intensity} shows the profiles of $A_{\mathrm{disp}}$ and $A'_{\mathrm{var}}$ aligned on this zero epoch. Both indices exhibit a systematic enhancement in the precursor window ($-96$ to $-24$~h), and the amplitude increases monotonically with storm intensity. For $A_{\mathrm{disp}}$, the precursor-window mean rises from $1.70$ (minor, $n=496$) to $1.95$ (moderate, $n=378$), $2.14$ (severe, $n=93$), and $2.53$ (extreme, $n=17$). For $A'_{\mathrm{var}}$, it rises from $0.38$ to $0.57$, $0.62$, and $0.99$. The monotonic scaling is consistent with stronger storms being driven by more energetic disturbances that produce stronger cosmic-ray modulation.\\

\subsubsection{Significance Against Random Epochs}\label{subsubsec:Epochs}

\begin{figure}[htbp]
\centering
\begin{minipage}{0.32\textwidth}
 \centering
 \includegraphics[width=\linewidth]{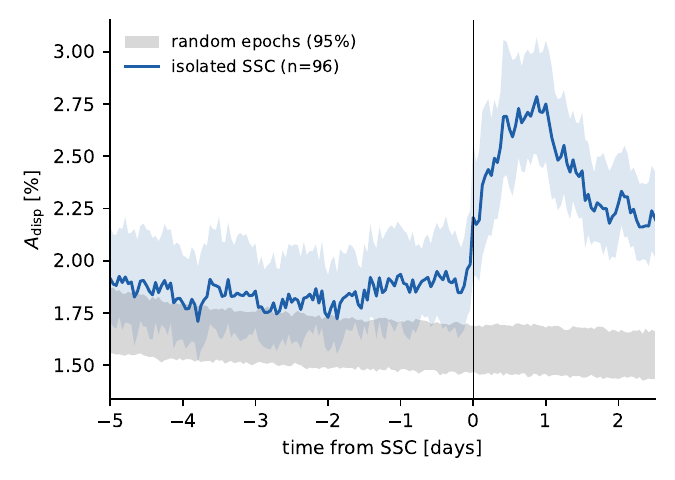}
 \textit{(a) \(A_{\mathrm{disp}}\)}
\end{minipage}\hfill
\begin{minipage}{0.32\textwidth}
 \centering
 \includegraphics[width=\linewidth]{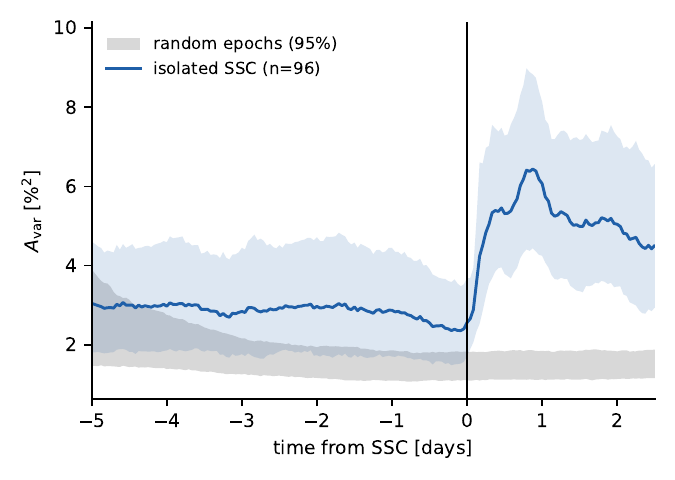}
 \textit{(b) \(A_{\mathrm{var}}\)}
\end{minipage}\hfill
\begin{minipage}{0.32\textwidth}
 \centering
 \includegraphics[width=\linewidth]{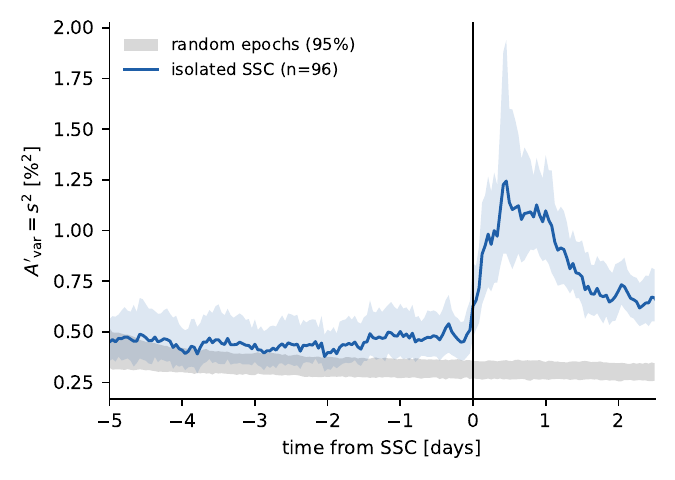}
 \textit{(c) \(A'_{\mathrm{var}}\)}
\end{minipage}
\caption{Superposed-epoch analysis of (a) \(A_{\mathrm{disp}}\), (b) \(A_{\mathrm{var}}\), and (c) \(A'_{\mathrm{var}}\) aligned on SSC for the 96 isolated events. The gray shaded region is the 95\% envelope from 500 random epochs drawn from the same calendar years.}
\label{fig:eps}
\end{figure}

To test whether this enhancement is statistically significant against background fluctuations, and to avoid the confounding effect of clustered storms, we restrict the random-epoch Monte Carlo analysis to isolated events. An event is defined as isolated if no other SSC or storm occurs within the preceding five days. This leaves 96 isolated events. For each event, a random epoch is drawn from the same calendar year, avoiding any storm or SSC within $\pm 5$~days, and the superposed analysis is repeated 500 times to obtain the 95\% envelope. Figure~\ref{fig:eps} shows the results. The \(A_{\mathrm{disp}}\) and \(A'_{\mathrm{var}}\) curves lie above the random envelope across most of the precursor window; \(A_{\mathrm{var}}\), which retains the network-mean term, is shown only as a total-modulation reference. The enhancement is broad; we do not define a lead time from the first crossing. Because the indices are built from a time-varying station subset, we explicitly test the effect of station selection using the fixed seven-station subset in Section~\ref{subsec:limits}.

\subsection{Principal-Component Analysis as a Diagnostic of the Global Mode}\label{subsec:pca_results}

\begin{figure}[htbp]
\centering
\begin{minipage}{0.48\textwidth}
 \centering
 \includegraphics[width=\linewidth]{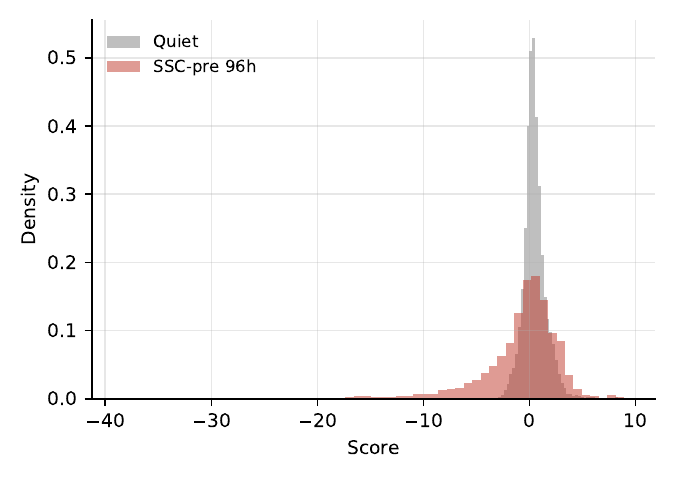}
 \textit{(a) PC1}
\end{minipage}\hfill
\begin{minipage}{0.48\textwidth}
 \centering
 \includegraphics[width=\linewidth]{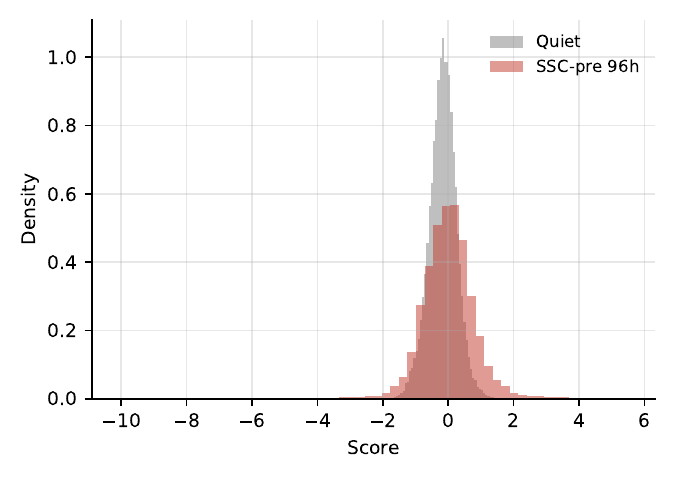}
 \textit{(b) PC2}
\end{minipage}
\caption{Distributions of the PC1 score (a) and PC2 score (b) during quiet periods (gray) and the 96~h pre-storm window (red). The histograms are density-normalized. Both distributions are unimodal and strongly overlapping: the red pre-storm distribution is a small shifted version of the gray quiet distribution, not a separate population.}
\label{fig:pca}
\end{figure}

The first two principal components explain 83.7\% and 5.5\% of the standardized residual variance, respectively. Their interpretation is not symmetric. PC1 has all-positive loadings (0.35--0.39) and is almost collinear with the network mean (Spearman $\rho=0.997$). It therefore represents a network-wide common mode: a lower PC1 score corresponds to a more negative $\bar{R}$, i.e., a global GCR depression. The PC1 shift is a shift of the global mode, not evidence for inter-station structure. PC2 has mixed-sign loadings (PTFM and TSMB positive; OULU, SOPO, THUL, and KERG negative; JUNG close to zero) and is only weakly correlated with the inter-station variance $A'_{\mathrm{var}}=s^2$ ($\rho=0.14$) and with $A_{\mathrm{disp}}$ ($\rho=0.13$). It is also only weakly correlated with the total network variance $A_{\mathrm{var}}$ ($\rho=0.08$). PC2 is therefore a weak secondary mode, not a strong proxy for inter-station dispersion.\\
Figure~\ref{fig:pca} shows the PC1 and PC2 distributions during quiet periods (gray) and the 96~h preceding SSC (red). Both distributions are unimodal and broad, and the red pre-storm distribution is a small shifted version of the gray quiet distribution rather than a separate population. PC1 shifts negative (Cliff's $\delta=-0.184$) and PC2 shifts positive ($\delta=+0.154$). A 96-h block permutation (5000 shuffles) gives $p=0.0002$ for both components. With 19,235 pre-storm hours and 33,054 quiet hours, even a small mean shift is resolvable; the effect sizes, however, are modest and the overlap remains large. The block test is essential: an hourly permutation would treat autocorrelated hours as independent and overstate the significance. The PCA thus detects a weak storm-related signal in the residuals, but it does not separate pre-storm from quiet hours at the event level. The PCA therefore provides a purely statistical, complementary indication that the multi-station residuals contain storm-related variance. It does not provide independent evidence for an inter-station precursor: PC1 is essentially the network mean, and PC2 is a weak secondary mode. This is consistent with the limited event-level warning skill in Section~\ref{subsec:alert}: the signal is detectable in aggregate but does not dominate the inter-station variance.

\subsection{Case Validation} \label{subsec:cases}
Do the statistically identified patterns manifest in individual storm events? We examine two representative cases: the 2003 November extreme storm and the 2018 August severe storm.\\

\subsubsection{2003 November 20 Extreme Storm}

\begin{figure}[h]
\centering
\includegraphics[width=\textwidth]{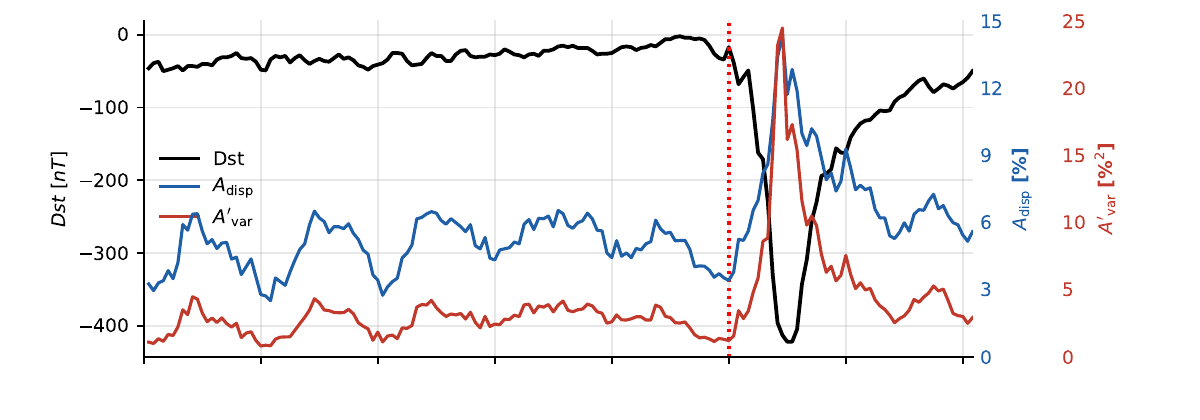}
\vspace{-0.5em}
\includegraphics[width=\textwidth]{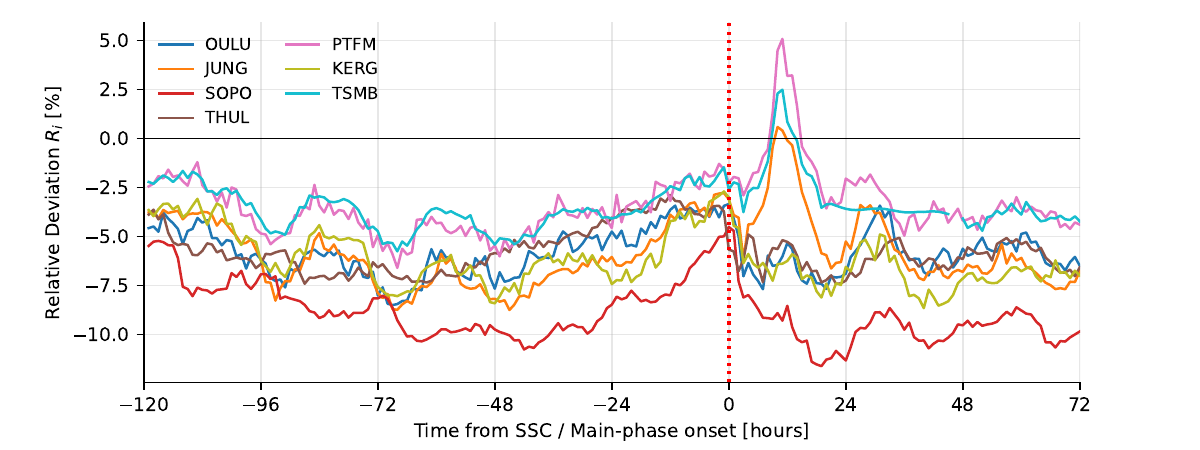}
\caption{Cosmic ray variations during the 2003 November 20 extreme geomagnetic storm (\(\mathrm{Dst_{min}} = -422\)~nT). \textit{Top panel}: Time series of the Dst index (black, left axis) alongside the two inter-station fluctuation indices: \(A_{\mathrm{disp}}\) (blue, right inner axis) and \(A'_{\mathrm{var}}\) (red, right outer axis). \textit{Bottom panel}: Time series of the relative cosmic ray flux deviations from quiet-period baselines for all seven neutron monitor stations (OULU, JUNG, SOPO, THUL, KERG, PTFM, TSMB). Each colored curve represents a different station. Note the initial synchronized decrease followed by spatially divergent responses as the storm main phase develops, which drives the pronounced enhancement of \(A_{\mathrm{disp}}\) and \(A'_{\mathrm{var}}\) shown in the top panel. The vertical red dotted line marks the SSC.}
\label{fig:case2003}
\end{figure}

The extreme GS of 2003 November 20 ($\mathrm{Dst_{min}} = -422$~nT), associated with the intense solar activity of solar cycle 23, provides an ideal test case (Figure~\ref{fig:case2003}). As shown in the bottom panel, prior to the SSC, the cosmic-ray flux at all stations drifted gradually toward negative deviations but remained relatively coherent across the network. However, within about 10~hours after the SSC (marked by the red dotted line), the cosmic-ray responses across stations diverged sharply, departing from the expectation of a uniform global Forbush decrease. While several stations (e.g., JUNG, PTFM) showed a rapid flux rebound, others (e.g., SOPO, KERG) continued to decline deeply. These differential responses reflected the complex interplay of each station's asymptotic viewing direction, geomagnetic cutoff rigidity, and the asymmetric magnetic structure of the CME flux rope.\\This spatially dependent differential response is explicitly captured in the top panel. As the Dst index plunged (downward black curve), the inter-station fluctuation indices $A_{\mathrm{disp}}$ (blue curve) and $A'_{\mathrm{var}}$ (red curve) exhibited a sharp upward surge. This diverse behavior originated from the asymmetric modulation by the CME magnetic flux rope and gradient-induced particle drifts. Because this complex spatial heterogeneity cannot be captured by single-station measurements or simple latitudinal/rigidity-based classifications, our direction-free fluctuation indices provide a quantitative and robust characterization of the disturbance.\\

\subsubsection{2018 August 26 Severe Storm}

\begin{figure}[h]
\centering
\includegraphics[width=\textwidth]{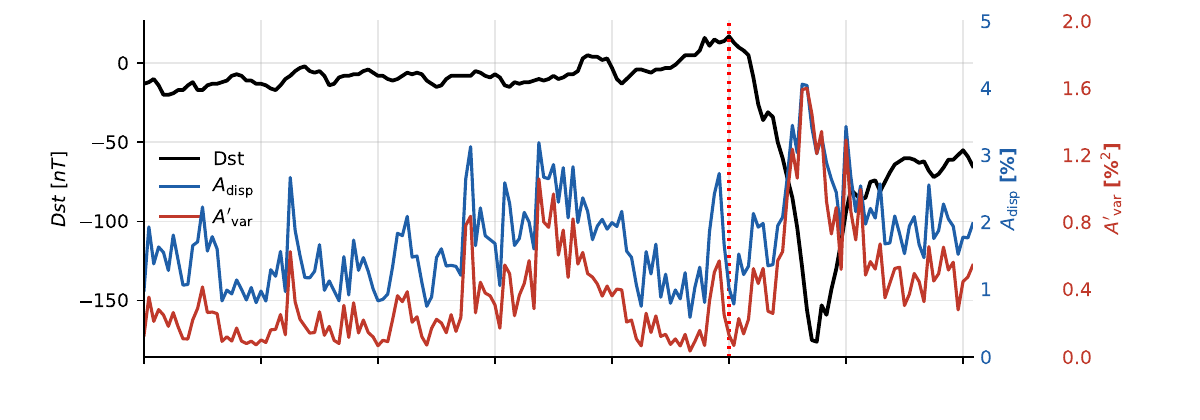}
\vspace{-0.5em}
\includegraphics[width=\textwidth]{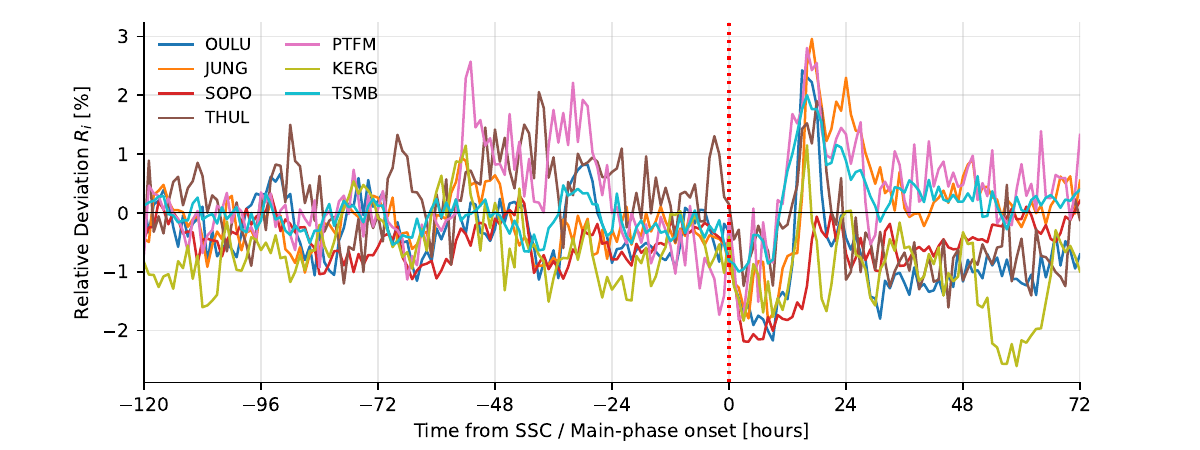}
\caption{Time series of the 2018 August 26 severe storm (\(\mathrm{Dst_{min}} = -176\)~nT). \textit{Top panel}: Dst index (black, left axis) alongside the inter-station fluctuation indices \(A_{\mathrm{disp}}\) (blue, right inner axis) and \(A'_{\mathrm{var}}\) (red, right outer axis). \textit{Bottom panel}: Relative cosmic-ray flux deviations \(R_i(t)\) for the seven stations. The red dotted vertical line marks the main-phase onset. Note the non-synchronous flux variations before the main-phase onset and the subsequent spatially divergent responses during the main phase.}
\label{fig:case2018}
\end{figure}

The 2018 August 26 severe storm (\(\mathrm{Dst_{min}} = -176\)~nT) exhibits a complex spatiotemporal evolution (Figure~\ref{fig:case2018}). In the bottom panel, during the 40--80~hours preceding the main-phase onset, prominent asynchronous cosmic-ray flux deviations emerge across the neutron-monitor network. For example, the PTFM station shows large positive excursions, while SOPO remains persistently suppressed to negative values. The network indices in the top panel remain variable during this interval and begin to rise in the final \(\sim\)20~hours before the steep Dst decrease, but the clearest enhancement occurs near the onset.\\
After the onset and before the Dst minimum, the station responses converge into a synchronized global decrease, corresponding to the classic Forbush decrease. Near the Dst minimum, the cosmic-ray intensity rebounds synchronously across stations. Following this rebound, the station responses diverge sharply once again: PTFM and TSMB exhibit rapid positive excursions, whereas OULU and KERG remain suppressed at low values. This evolutionary sequence---asynchronous pre-onset deviations, a brief interval of coordinated global suppression after the onset, a synchronous rebound near the Dst trough, and subsequent divergent recovery---illustrates the spatially variable nature of the GCR perturbation driven by the CME and demonstrates the sensitivity of our direction-free fluctuation indices to inter-station heterogeneity.\\
Side-by-side examination of the 2003 and 2018 storms illustrates contrasting evolutionary behaviors, consistent with differences in their CME configurations. The 2003 extreme event is characterized by a relatively quiet pre-SSC period, where inter-station divergence emerges promptly following shock passage. The 2018 severe storm behaves differently: prominent asynchronous fluctuations appear well ahead of the main-phase onset, followed by a synchronized global Forbush decrease and divergent recovery at later stages. Despite such differences in their temporal evolution, both events develop measurable inter-station heterogeneity that is directly quantified by \(A_{\mathrm{disp}}\) and \(A'_{\mathrm{var}}\). This supports the utility of our direction-free fluctuation indices as robust diagnostics for spatially non-uniform GCR perturbations across different types of CME-driven geomagnetic storms.\\

\subsection{Intensity Correlation and Event-Level Alert Skill}
\label{subsec:alert}

\begin{figure}[h]
\centering
\includegraphics[width=0.85\textwidth]{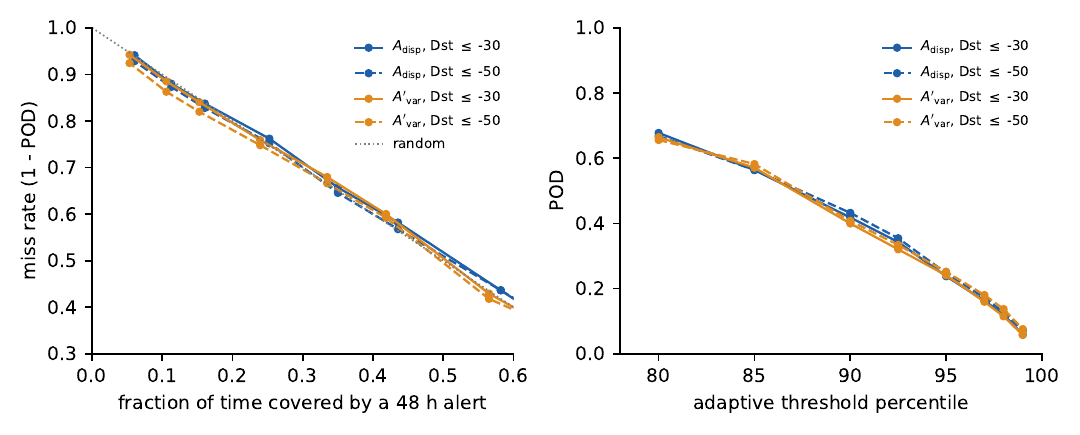}
\caption{Warning skill assessment. Left: Molchan diagram for \(A_{\mathrm{disp}}\) and \(A'_{\mathrm{var}}\) at different threshold percentiles. The dashed diagonal line represents random alerts. Right: Probability of detection (POD) as a function of the adaptive threshold percentile.}
\label{fig:molchan}
\end{figure}

\begin{figure}[h]
\centering
\includegraphics[width=0.55\textwidth]{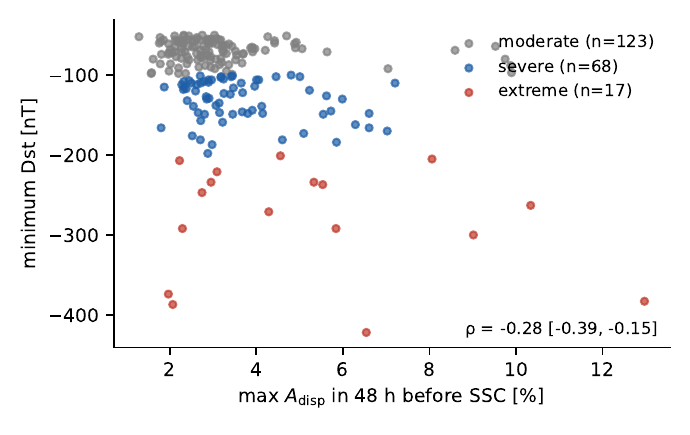}
\caption{Maximum \(A_{\mathrm{disp}}\) in the 48~h before SSC versus the minimum Dst for all storm-producing events. Gray, blue, and red dots correspond to moderate (\(n=123\)), severe (\(n=68\)), and extreme (\(n=17\)) storms, respectively.}
\label{fig:intensity}
\end{figure}

The event-level alert evaluation uses 983 storm onsets and 1,243 ($A_{\mathrm{disp}}$) or 1,165 ($A'_{\mathrm{var}}$) alert episodes. At the 95th-percentile threshold, $A_{\mathrm{disp}}$ achieves POD $=23.8\%$, FAR $=73.4\%$, and coverage $C=25.3\%$; $A'_{\mathrm{var}}$ achieves POD $=24.1\%$, FAR $=72.0\%$, and $C=23.9\%$. The Molchan skill is $S=-0.010$ for $A_{\mathrm{disp}}$ and $S=-0.004$ for $A'_{\mathrm{var}}$, i.e., there is no positive skill (Figure~\ref{fig:molchan}). For the $\mathrm{Dst}\leq-50$~nT subset, the POD is similar (24.8\% and 25.2\%) but the FAR is higher (85.7\% and 85.0\%); lowering the threshold increases the POD but also the coverage and the FAR, and no operating point yields positive Molchan skill.
Within the 208 storms with a cataloged SSC and $\mathrm{Dst}_{\min}\leq-50$~nT, the maximum $A_{\mathrm{disp}}$ in the 48~h before SSC correlates with $\mathrm{Dst}_{\min}$ at Spearman $\rho=-0.28$ (95\% CI $-0.39$ to $-0.15$; $n=208$; Figure~\ref{fig:intensity}). The median increases from 2.85 (moderate, $n=123$) to 3.20 (severe, $n=68$) and 4.56 (extreme, $n=17$), but the substantial overlap between intensity classes limits the index to descriptive characterization rather than operational intensity forecasting. The AUC for separating severe from moderate storms is 0.66 (95\% CI 0.58--0.74). Using $A'_{\mathrm{var}}$ instead gives $\rho=-0.27$ and AUC $=0.66$.

\section{Discussion} \label{sec:discussion}

\subsection{Physical Interpretation} \label{subsec:physics}
The observed pre-storm enhancement of the fluctuation indices reflects the shared interplanetary driver. CMEs and their driven shocks modulate GCRs through magnetic field enhancement and turbulent scattering, while simultaneously transferring energy to Earth's magnetosphere through magnetic reconnection \cite{Cane2000,Gonzalez1994,Schwenn2006}. The monotonic scaling of precursor amplitude with storm intensity (Section~\ref{subsubsec:Precursor}) may arise because more energetic CMEs produce both stronger GCR modulation and more intense geomagnetic disturbances \cite{Mishra2008}.\\
The inter-station fluctuation indices $A_{\mathrm{disp}}$ and \(A'_{\mathrm{var}}\) respond to the spatial non-uniformity of this modulation: as the CME's extended magnetic structure approaches, different stations, with their distinct asymptotic viewing directions and cutoff rigidities, sample the disturbance differently.\\The simultaneous plunge in the Dst index and the surge in the fluctuation indices $A_{\mathrm{disp}}$ and \(A'_{\mathrm{var}}\) during the main phase (Figures~\ref{fig:case2003} and \ref{fig:case2018}) is consistent with this picture. The CME-driven shock and sheath compress the heliospheric magnetic field, producing a global Forbush decrease, while the asymmetric flux-rope structure and gradient-induced particle drifts cause the station responses to diverge. A critical factor in this spatial divergence is the geomagnetic cutoff rigidity, which varies significantly across our network (from \(\sim 0.1\)~GV at SOPO to \(\sim 9.15\)~GV at TSMB). Low-rigidity stations are highly sensitive to low-energy cosmic rays, which are strongly suppressed by the approaching magnetic barrier. In contrast, high-rigidity stations primarily record high-energy particles, which are less affected by the interplanetary turbulence and can even exhibit transient enhancements driven by drift effects. This rigidity-dependent response, combined with differing asymptotic viewing directions, produces the complex, non-uniform station behavior observed in the case studies. The resulting spatial heterogeneity is directly quantified by our direction-free indices, without requiring spherical harmonic reconstruction or prior knowledge of the IMF orientation.\\

\subsection{Forecasting Performance Limitations}
\label{subsec:Limitations}

The null alert skill is not an artifact arising from the causal baseline or SSC alignment: the pre-storm enhancement is broad and slowly varying, so a simple threshold cannot reliably separate pre-storm windows from other disturbed periods without incurring a high false-alarm rate. One plausible interpretation is that our direction-free indices respond mainly to the large-scale shock and magnetic-barrier structure of CMEs, whereas the resulting geomagnetic storm strength is governed primarily by the southward IMF component ($B_z$). Since our indices do not use in-situ solar-wind magnetic data, they lack direct sensitivity to the embedded $B_z$ that controls storm severity. This picture is consistent with the weak-to-moderate intensity correlation reported above, but it is not definitive; dedicated tests using solar-wind data and direction-resolved cosmic-ray observables are needed. Therefore, robust event-level forecasting will require complementary information, such as solar-wind observations or IMF $B_z$ constraints, rather than improvements to cosmic-ray diagnostics in isolation.\\

\subsection{Physical Basis for Time Window Selection} \label{subsec:timewindow}

The 96-h (four-day) window is grounded in CME propagation physics. Fast CMEs with initial speeds above 1000~km~s$^{-1}$ typically reach Earth within 1--2~days, while slower CMEs ($\sim$400--600~km~s$^{-1}$) or those with significant deflection from the Sun--Earth line may require 4--5~days \cite{Gopalswamy2001,Manoharan2004}. Statistical and machine-learning studies of CME transit times give a median Sun-to-Earth transit of approximately 80~h, with a broad distribution spanning 20--120~h \cite{Gopalswamy2001,Vrsnak2013,Zhou2026ApJ}. A 96-h window is long enough to cover slow CMEs and short enough to limit contamination from unrelated solar activity.\\For GCR-based precursor analysis, the relevant timescale is not the total CME transit time but the interval over which the CME's extended magnetic structure begins to modulate GCRs detectable at Earth. The sheath and shock region ahead of the ejecta can influence GCR propagation before the main body arrives, producing inter-station fluctuations over tens of hours. The 96-h window encompasses this interval. It is an analysis interval, not a lead time: the results in Section~\ref{subsubsec:Epochs} show a broad enhancement across the window, not a sharp onset at a well-defined number of hours before the storm.\\CIR/HSS-driven disturbances follow different timescales. The peak of the associated high-speed stream is typically observed at Earth about three days after the coronal-hole area maximizes near the central meridian \cite{Vrsnak2007}. CIR-associated GCR modulation is weaker and more gradual, so these events are less suitable for GCR-based precursor analysis. Because our sample includes both driver types, the measured ensemble signal is likely a mixture; the CME-dominated severe and extreme classes are the cleanest part of the sample.

\subsection{Limitations and Future Directions} \label{subsec:limits}

Several limitations remain. First, the current NM network exhibits spatial asymmetry: SOPO lacks data from 2006 to 2009, and PTFM from 2012 to 2015, so the effective network size varies over the study period. Hours with valid data from fewer than four stations are excluded, but the non-uniform coverage may still introduce residual biases. The fixed seven-station subset provides a direct test of station-selection effects. Restricting to hours with all seven stations leaves 109,838 of 219,144 hours (50.1\%) and 63 of the 96 isolated events with at least one valid pre-storm hour. The window-mean enhancement remains significant for $A'_{\mathrm{var}}$ ($p=0.004$) but weakens for $A_{\mathrm{disp}}$ ($p=0.012$); the latter is a max--min statistic and is therefore more sensitive to which stations are present. We therefore treat $A'_{\mathrm{var}}$ as the more robust index and $A_{\mathrm{disp}}$ as descriptive. The monotonic intensity ordering is preserved in both cases.\\Second, the fluctuation indices $A_{\mathrm{disp}}$ and \(A'_{\mathrm{var}}\) treat all stations with equal weight, without applying a rigidity-dependent coupling coefficient. Because low-rigidity stations are more susceptible to low-energy cosmic-ray modulation, their deviations might disproportionately drive the network variance during certain storm phases. While this equal-weight scheme is computationally efficient and preserves the direction-free property of our indices, future work could incorporate station-specific yield functions and cutoff rigidities to refine index sensitivity.\\Third, the indices capture only the total inter-station dispersion and do not resolve the direction of the anisotropy. A full spherical-harmonic reconstruction requires calibrated asymptotic viewing directions and a larger network (9--15 stations), which is beyond the scope of the present seven-station analysis. A muon network with pitch-angle resolution would provide a direct test of whether the pre-storm dispersion is dominated by field-aligned streaming, loss-cone effects, or gradient-induced drifts.\\Fourth, the SSC catalog covers only 29\% of the identified storms, which limits the sample available for SSC-aligned analysis. Storms without a cataloged SSC are aligned on the main-phase onset, introducing a small timing uncertainty. Continued accumulation of SSC events would improve this.\\Finally, extreme events remain challenging due to sample scarcity. Our dataset contains only 17 extreme storms. This small sample size directly limits the statistical confidence for this most consequential category, as evidenced by the visibly wider bootstrap confidence intervals for the extreme class in Figure~\ref{fig:sea_intensity}. Retrospective analysis of historical extreme events and continued data accumulation will be essential to further refine the precursor characteristics of these rare but highly impactful events.\\

\section{Conclusions} \label{sec:conclusions}

We have analyzed 25 years of neutron-monitor data from seven stations (1995--2019) to characterize the relationship between inter-station cosmic-ray fluctuations and geomagnetic storms. Our two main conclusions are as follows.\\
First, both indices are enhanced, on average, in the days before storm onset, and the window-mean amplitude increases monotonically with storm intensity. For isolated events, the enhancement is significant against a quiet random control ($p=0.002$), but it is broad and gradual: the \(A_{\mathrm{disp}}\) and \(A'_{\mathrm{var}}\) curves lie above the envelope across the \(-96\) to \(-24\)~h window. We therefore describe it as an average pre-storm enhancement, not as an event-specific lead time. This result is built on a strictly causal baseline and two direction-free indices; a variance decomposition shows that the raw network variance is dominated by the network-mean term, so $A'_{\mathrm{var}}$ is used to isolate the inter-station component. The enhancement is robust to station-count normalization: the $1/n$ definition and the seven-station subset preserve the intensity ordering; the seven-station significance is weaker for $A_{\mathrm{disp}}$ ($p\approx0.012$) than for $A'_{\mathrm{var}}$ ($p\approx0.004$).\\
Second, the indices do not provide an operational event-level warning. The fluctuation indices correlate with storm intensity at Spearman $\rho=-0.28$ (AUC $0.66$), but the overlap between intensity classes is too large for intensity forecasting. A principal-component analysis shows that the leading mode (83.7\%) is almost collinear with the network mean ($\rho=0.997$) and therefore represents a global GCR depression, not inter-station structure; the second mode (5.5\%) is weak and only weakly correlated with $A_{\mathrm{disp}}$ ($\rho=0.13$). A threshold-based alert evaluated with a Molchan diagram does not exceed the random-alert baseline at the event level (POD $=23.8\%$ and $24.1\%$; FAR $=73.4\%$ and $72.0\%$ for $A_{\mathrm{disp}}$ and $A'_{\mathrm{var}}$ at the 95th-percentile threshold).\\
The inter-station fluctuation indices are therefore a descriptive precursor of the disturbed heliospheric state, not an operational intensity forecaster. Their value lies in providing a direction-free, causally computable characterization of the spatial non-uniformity of GCR modulation. A reliable event-level warning will require combining these indices with independent information---solar-wind measurements, southward IMF estimates, or another cosmic-ray observable---rather than refining the cosmic-ray analysis alone. Event-specific lead times and driver-dependent performance are separate questions that require a dedicated event-level study.

\section*{Data Availability Statement}

The neutron-monitor data are available from the Neutron Monitor Database (NMDB; https://www.nmdb.eu) and the individual station PI teams. The hourly Dst index is from the World Data Center for Geomagnetism, Kyoto (https://wdc.kugi.kyoto-u.ac.jp). The SSC catalog is from the Ebre Observatory (https://www.obsebre.es). The derived hourly indices, storm and SSC catalogs, PCA scores, and analysis code are archived at Zenodo under DOI 10.5281/zenodo.23971667 (MIT license). No new raw neutron-monitor data were created.\\
Software/Code Availability: The analysis code (Python 3.10) is available at GitHub (\url{https://github.com/LHY-lhy-050909/cosmic_ray_analysis_repo}) and archived at Zenodo under DOI 10.5281/zenodo.23971667 (MIT license). It includes the causal baseline, index definitions, superposed-epoch and random-epoch analysis, PCA block permutation, event-level alert evaluation, and figure scripts.

\section*{Author Contributions}

{Zongyuan Ge: Conceptualization, Formal analysis, Investigation, Methodology, Writing--original draft.
Haoyang Li: Data curation, Formal analysis, Investigation, Methodology, Writing--original draft.
Zhaoming Wang: Supervision, Writing--review \& editing.}

\authorcontributions
\bibliography{re}

\begin{thebibliography}{}

\bibitem [\protect \citeauthoryear {%
{Bala}%
\ \BBA {} {Reiff}%
}{%
{Bala}%
\ \BBA {} {Reiff}%
}{%
{\protect \APACyear {2012}}%
}]{%
Bala2012}
\APACinsertmetastar {%
Bala2012}%
\begin{APACrefauthors}%
{Bala}, R.%
\BCBT {}\ \BBA {} {Reiff}, P.%
\end{APACrefauthors}%
\unskip\
\newblock
\APACrefYearMonthDay{2012}{}{}.
\newblock
{\BBOQ}\APACrefatitle {{Improvements in short-term forecasting of geomagnetic
  activity}} {{Improvements in short-term forecasting of geomagnetic
  activity}}.{\BBCQ}
\newblock
\APACjournalVolNumPages{Space Weather}{10}{6}{S06001}.
\newblock
\begin{APACrefDOI} \doi{10.1029/2012SW000779} \end{APACrefDOI}
\PrintBackRefs{\CurrentBib}

\bibitem [\protect \citeauthoryear {%
{Bell}%
}{%
{Bell}%
}{%
{\protect \APACyear {1978}}%
}]{%
Bell1978}
\APACinsertmetastar {%
Bell1978}%
\begin{APACrefauthors}%
{Bell}, A\BPBI R.%
\end{APACrefauthors}%
\unskip\
\newblock
\APACrefYearMonthDay{1978}{}{}.
\newblock
{\BBOQ}\APACrefatitle {{The acceleration of cosmic rays in shock fronts -- I}}
  {{The acceleration of cosmic rays in shock fronts -- I}}.{\BBCQ}
\newblock
\APACjournalVolNumPages{Monthly Notices of the Royal Astronomical
  Society}{182}{}{147-156}.
\newblock
\begin{APACrefDOI} \doi{10.1093/mnras/182.2.147} \end{APACrefDOI}
\PrintBackRefs{\CurrentBib}

\bibitem [\protect \citeauthoryear {%
Belov%
\ \protect \BOthers {.}}{%
Belov%
\ \protect \BOthers {.}}{%
{\protect \APACyear {1995}}%
}]{%
Belov1995}
\APACinsertmetastar {%
Belov1995}%
\begin{APACrefauthors}%
Belov, A\BPBI V.%
, Dorman, L\BPBI I.%
, Eroshenko, E\BPBI A.%
, Iucci, N.%
, Villoresi, G.%
\BCBL {}\ \BBA {} Yanke, V\BPBI G.%
\end{APACrefauthors}%
\unskip\
\newblock
\APACrefYearMonthDay{1995}{08}{}.
\newblock
{\BBOQ}\APACrefatitle {Search for Predictors of Forbush Decreases} {Search for
  predictors of forbush decreases}.{\BBCQ}
\newblock
\BIn{} \APACrefbtitle {24th International Cosmic Ray Conference} {24th
  international cosmic ray conference}\ (\BVOL~4, \BPGS\ 888--891).
\PrintBackRefs{\CurrentBib}

\bibitem [\protect \citeauthoryear {%
Belov%
, Eroshenko%
, Oleneva%
\BCBL {}\ \BBA {} Yanke%
}{%
Belov%
\ \protect \BOthers {.}}{%
{\protect \APACyear {2009}}%
}]{%
Belov2009}
\APACinsertmetastar {%
Belov2009}%
\begin{APACrefauthors}%
Belov, A\BPBI V.%
, Eroshenko, E\BPBI A.%
, Oleneva, V\BPBI A.%
\BCBL {}\ \BBA {} Yanke, V\BPBI G.%
\end{APACrefauthors}%
\unskip\
\newblock
\APACrefYearMonthDay{2009}{}{}.
\newblock
{\BBOQ}\APACrefatitle {Ground level enhancements of the solar cosmic rays and
  Forbush decreases in 23rd solar cycle} {Ground level enhancements of the
  solar cosmic rays and forbush decreases in 23rd solar cycle}.{\BBCQ}
\newblock
\BIn{} \APACrefbtitle {Proceedings of the 31st International Cosmic Ray
  Conference.} {Proceedings of the 31st international cosmic ray conference.}
\newblock
\APACaddressPublisher{Łódź, Poland}{}.
\newblock
\APACrefnote{Paper 1004}
\PrintBackRefs{\CurrentBib}

\bibitem [\protect \citeauthoryear {%
{Bieber}%
\ \BBA {} {Evenson}%
}{%
{Bieber}%
\ \BBA {} {Evenson}%
}{%
{\protect \APACyear {1998}}%
}]{%
Bieber1998}
\APACinsertmetastar {%
Bieber1998}%
\begin{APACrefauthors}%
{Bieber}, J\BPBI W.%
\BCBT {}\ \BBA {} {Evenson}, P.%
\end{APACrefauthors}%
\unskip\
\newblock
\APACrefYearMonthDay{1998}{}{}.
\newblock
{\BBOQ}\APACrefatitle {{CME geometry in relation to cosmic ray anisotropy}}
  {{CME geometry in relation to cosmic ray anisotropy}}.{\BBCQ}
\newblock
\APACjournalVolNumPages{Geophysical Research Letters}{25}{15}{2955-2958}.
\newblock
\begin{APACrefDOI} \doi{10.1029/98GL51232} \end{APACrefDOI}
\PrintBackRefs{\CurrentBib}

\bibitem [\protect \citeauthoryear {%
{Cane}%
}{%
{Cane}%
}{%
{\protect \APACyear {2000}}%
}]{%
Cane2000}
\APACinsertmetastar {%
Cane2000}%
\begin{APACrefauthors}%
{Cane}, H\BPBI V.%
\end{APACrefauthors}%
\unskip\
\newblock
\APACrefYearMonthDay{2000}{}{}.
\newblock
{\BBOQ}\APACrefatitle {{Coronal mass ejections and Forbush decreases}}
  {{Coronal mass ejections and Forbush decreases}}.{\BBCQ}
\newblock
\APACjournalVolNumPages{Space Science Reviews}{93}{1-2}{55-77}.
\newblock
\begin{APACrefDOI} \doi{10.1023/A:1026532125747} \end{APACrefDOI}
\PrintBackRefs{\CurrentBib}

\bibitem [\protect \citeauthoryear {%
{Clem}%
\ \BBA {} {Dorman}%
}{%
{Clem}%
\ \BBA {} {Dorman}%
}{%
{\protect \APACyear {2000}}%
}]{%
Clem2000}
\APACinsertmetastar {%
Clem2000}%
\begin{APACrefauthors}%
{Clem}, J\BPBI M.%
\BCBT {}\ \BBA {} {Dorman}, L\BPBI I.%
\end{APACrefauthors}%
\unskip\
\newblock
\APACrefYearMonthDay{2000}{}{}.
\newblock
{\BBOQ}\APACrefatitle {{Neutron monitor response functions}} {{Neutron monitor
  response functions}}.{\BBCQ}
\newblock
\APACjournalVolNumPages{Space Science Reviews}{93}{1-2}{335-359}.
\newblock
\begin{APACrefDOI} \doi{10.1023/A:1026508915269} \end{APACrefDOI}
\PrintBackRefs{\CurrentBib}

\bibitem [\protect \citeauthoryear {%
{Duggal}%
\ \BBA {} {Pomerantz}%
}{%
{Duggal}%
\ \BBA {} {Pomerantz}%
}{%
{\protect \APACyear {1976}}%
}]{%
Duggal1976}
\APACinsertmetastar {%
Duggal1976}%
\begin{APACrefauthors}%
{Duggal}, S\BPBI P.%
\BCBT {}\ \BBA {} {Pomerantz}, M\BPBI A.%
\end{APACrefauthors}%
\unskip\
\newblock
\APACrefYearMonthDay{1976}{}{}.
\newblock
{\BBOQ}\APACrefatitle {{Origin of transient north-south anisotropy of cosmic
  rays}} {{Origin of transient north-south anisotropy of cosmic rays}}.{\BBCQ}
\newblock
\APACjournalVolNumPages{Journal of Geophysical Research}{81}{28}{5032-5038}.
\newblock
\begin{APACrefDOI} \doi{10.1029/JA081i028p05032} \end{APACrefDOI}
\PrintBackRefs{\CurrentBib}

\bibitem [\protect \citeauthoryear {%
{Dumbovi{\'c}}%
, {Vr{\v{s}}nak}%
, {{\v{C}}alogovi{\'c}}%
\BCBL {}\ \BBA {} {Karlica}%
}{%
{Dumbovi{\'c}}%
\ \protect \BOthers {.}}{%
{\protect \APACyear {2011}}%
}]{%
Dumbovic2011}
\APACinsertmetastar {%
Dumbovic2011}%
\begin{APACrefauthors}%
{Dumbovi{\'c}}, M.%
, {Vr{\v{s}}nak}, B.%
, {{\v{C}}alogovi{\'c}}, J.%
\BCBL {}\ \BBA {} {Karlica}, M.%
\end{APACrefauthors}%
\unskip\
\newblock
\APACrefYearMonthDay{2011}{}{}.
\newblock
{\BBOQ}\APACrefatitle {{Cosmic ray modulation by solar wind disturbances}}
  {{Cosmic ray modulation by solar wind disturbances}}.{\BBCQ}
\newblock
\APACjournalVolNumPages{Astronomy \& Astrophysics}{531}{}{A91}.
\newblock
\begin{APACrefDOI} \doi{10.1051/0004-6361/201016006} \end{APACrefDOI}
\PrintBackRefs{\CurrentBib}

\bibitem [\protect \citeauthoryear {%
{Forbush}%
}{%
{Forbush}%
}{%
{\protect \APACyear {1938}}%
}]{%
Forbush1938}
\APACinsertmetastar {%
Forbush1938}%
\begin{APACrefauthors}%
{Forbush}, S\BPBI E.%
\end{APACrefauthors}%
\unskip\
\newblock
\APACrefYearMonthDay{1938}{}{}.
\newblock
{\BBOQ}\APACrefatitle {{On world-wide changes in cosmic-ray intensity}} {{On
  world-wide changes in cosmic-ray intensity}}.{\BBCQ}
\newblock
\APACjournalVolNumPages{Physical Review}{54}{12}{975-988}.
\newblock
\begin{APACrefDOI} \doi{10.1103/PhysRev.54.975} \end{APACrefDOI}
\PrintBackRefs{\CurrentBib}

\bibitem [\protect \citeauthoryear {%
{Forman}%
\ \BBA {} {Gleeson}%
}{%
{Forman}%
\ \BBA {} {Gleeson}%
}{%
{\protect \APACyear {1975}}%
}]{%
Forman1975}
\APACinsertmetastar {%
Forman1975}%
\begin{APACrefauthors}%
{Forman}, M\BPBI A.%
\BCBT {}\ \BBA {} {Gleeson}, L\BPBI J.%
\end{APACrefauthors}%
\unskip\
\newblock
\APACrefYearMonthDay{1975}{}{}.
\newblock
{\BBOQ}\APACrefatitle {{Cosmic-ray streaming and anisotropies}} {{Cosmic-ray
  streaming and anisotropies}}.{\BBCQ}
\newblock
\APACjournalVolNumPages{Astrophysics and Space Science}{32}{}{77-94}.
\newblock
\begin{APACrefDOI} \doi{10.1007/BF00646218} \end{APACrefDOI}
\PrintBackRefs{\CurrentBib}

\bibitem [\protect \citeauthoryear {%
{Ge}%
, {Zhang}%
, {Zhou}%
, {Zeng}%
\BCBL {}\ \BBA {} {Zhou}%
}{%
{Ge}%
\ \protect \BOthers {.}}{%
{\protect \APACyear {2026}}%
}]{%
Ge2026}
\APACinsertmetastar {%
Ge2026}%
\begin{APACrefauthors}%
{Ge}, Z.%
, {Zhang}, C.%
, {Zhou}, W.%
, {Zeng}, H.%
\BCBL {}\ \BBA {} {Zhou}, G.%
\end{APACrefauthors}%
\unskip\
\newblock
\APACrefYearMonthDay{2026}{}{}.
\newblock
{\BBOQ}\APACrefatitle {{Physics-informed LSTM forecasting of geomagnetic storms
  using galactic cosmic-ray modulation signals}} {{Physics-informed LSTM
  forecasting of geomagnetic storms using galactic cosmic-ray modulation
  signals}}.{\BBCQ}
\newblock
\APACjournalVolNumPages{Research in Astronomy and Astrophysics}{26}{}{095021}.
\newblock
\begin{APACrefDOI} \doi{10.1088/1674-4527/ae61ec} \end{APACrefDOI}
\PrintBackRefs{\CurrentBib}

\bibitem [\protect \citeauthoryear {%
{Gonzalez}%
\ \protect \BOthers {.}}{%
{Gonzalez}%
\ \protect \BOthers {.}}{%
{\protect \APACyear {1994}}%
}]{%
Gonzalez1994}
\APACinsertmetastar {%
Gonzalez1994}%
\begin{APACrefauthors}%
{Gonzalez}, W\BPBI D.%
, {Joselyn}, J\BPBI A.%
, {Kamide}, Y.%
, {Kroehl}, H\BPBI W.%
, {Rostoker}, G.%
, {Tsurutani}, B\BPBI T.%
\BCBL {}\ \BBA {} {Vasyliunas}, V\BPBI M.%
\end{APACrefauthors}%
\unskip\
\newblock
\APACrefYearMonthDay{1994}{}{}.
\newblock
{\BBOQ}\APACrefatitle {{What is a geomagnetic storm?}} {{What is a geomagnetic
  storm?}}{\BBCQ}
\newblock
\APACjournalVolNumPages{Journal of Geophysical Research}{99}{A4}{5771-5792}.
\newblock
\begin{APACrefDOI} \doi{10.1029/93JA02867} \end{APACrefDOI}
\PrintBackRefs{\CurrentBib}

\bibitem [\protect \citeauthoryear {%
{Gopalswamy}%
, {Lara}%
, {Yashiro}%
, {Kaiser}%
\BCBL {}\ \BBA {} {Howard}%
}{%
{Gopalswamy}%
\ \protect \BOthers {.}}{%
{\protect \APACyear {2001}}%
}]{%
Gopalswamy2001}
\APACinsertmetastar {%
Gopalswamy2001}%
\begin{APACrefauthors}%
{Gopalswamy}, N.%
, {Lara}, A.%
, {Yashiro}, S.%
, {Kaiser}, M\BPBI L.%
\BCBL {}\ \BBA {} {Howard}, R\BPBI A.%
\end{APACrefauthors}%
\unskip\
\newblock
\APACrefYearMonthDay{2001}{}{}.
\newblock
{\BBOQ}\APACrefatitle {{Predicting the 1-AU arrival times of coronal mass
  ejections}} {{Predicting the 1-AU arrival times of coronal mass
  ejections}}.{\BBCQ}
\newblock
\APACjournalVolNumPages{Journal of Geophysical
  Research}{106}{A12}{29207-29217}.
\newblock
\begin{APACrefDOI} \doi{10.1029/2001JA000177} \end{APACrefDOI}
\PrintBackRefs{\CurrentBib}

\bibitem [\protect \citeauthoryear {%
{Grigoryev}%
, {Starodubtsev}%
\BCBL {}\ \BBA {} {Gololobov}%
}{%
{Grigoryev}%
\ \protect \BOthers {.}}{%
{\protect \APACyear {2017}}%
}]{%
Grigoryev2017}
\APACinsertmetastar {%
Grigoryev2017}%
\begin{APACrefauthors}%
{Grigoryev}, V\BPBI G.%
, {Starodubtsev}, S\BPBI A.%
\BCBL {}\ \BBA {} {Gololobov}, P\BPBI Y.%
\end{APACrefauthors}%
\unskip\
\newblock
\APACrefYearMonthDay{2017}{}{}.
\newblock
{\BBOQ}\APACrefatitle {{Monitoring geomagnetic disturbance predictors using
  data of ground measurements of cosmic rays}} {{Monitoring geomagnetic
  disturbance predictors using data of ground measurements of cosmic
  rays}}.{\BBCQ}
\newblock
\APACjournalVolNumPages{Bulletin of the Russian Academy of Sciences:
  Physics}{81}{2}{200-202}.
\newblock
\begin{APACrefDOI} \doi{10.3103/S1062873817020198} \end{APACrefDOI}
\PrintBackRefs{\CurrentBib}

\bibitem [\protect \citeauthoryear {%
{Idosa}%
, {Giri}%
, {Adhikari}%
, {Mosisa}%
\BCBL {}\ \BBA {} {Gashu}%
}{%
{Idosa}%
\ \protect \BOthers {.}}{%
{\protect \APACyear {2023}}%
}]{%
Idosa2023}
\APACinsertmetastar {%
Idosa2023}%
\begin{APACrefauthors}%
{Idosa}, C.%
, {Giri}, A.%
, {Adhikari}, B.%
, {Mosisa}, E.%
\BCBL {}\ \BBA {} {Gashu}, C.%
\end{APACrefauthors}%
\unskip\
\newblock
\APACrefYearMonthDay{2023}{}{}.
\newblock
{\BBOQ}\APACrefatitle {Variations of cosmic ray intensity with the solar flare
  index, coronal index, and geomagnetic indices: Wavelet and cross correlation
  approaches} {Variations of cosmic ray intensity with the solar flare index,
  coronal index, and geomagnetic indices: Wavelet and cross correlation
  approaches}.{\BBCQ}
\newblock
\APACjournalVolNumPages{Physics of Plasmas}{30}{8}{083901}.
\newblock
\begin{APACrefDOI} \doi{10.1063/5.0157553} \end{APACrefDOI}
\PrintBackRefs{\CurrentBib}

\bibitem [\protect \citeauthoryear {%
{Kojima}%
\ \protect \BOthers {.}}{%
{Kojima}%
\ \protect \BOthers {.}}{%
{\protect \APACyear {2025}}%
}]{%
Kojima2025}
\APACinsertmetastar {%
Kojima2025}%
\begin{APACrefauthors}%
{Kojima}, H.%
, {Shibata}, S.%
, {Oshima}, A.%
, {Takamaru}, H.%
, {Koi}, T.%
, {Yamazaki}, K.%
\BDBL {}{Arunbabu}, K.%
\end{APACrefauthors}%
\unskip\
\newblock
\APACrefYearMonthDay{2025}{}{}.
\newblock
{\BBOQ}\APACrefatitle {Response of Galactic Cosmic Rays to Solar Disturbances:
  A Study of the May 2024 Forbush Decrease with GRAPES-3 Muon Telescope}
  {Response of galactic cosmic rays to solar disturbances: A study of the may
  2024 forbush decrease with grapes-3 muon telescope}.{\BBCQ}
\newblock
\BIn{} \APACrefbtitle {Proceedings of Science (PoS), 39th International Cosmic
  Ray Conference (ICRC2025)} {Proceedings of science (pos), 39th international
  cosmic ray conference (icrc2025)}\ (\BVOL\ ICRC2025, \BPG~1310).
\newblock
\begin{APACrefDOI} \doi{10.22323/1.501.1310} \end{APACrefDOI}
\PrintBackRefs{\CurrentBib}

\bibitem [\protect \citeauthoryear {%
Loto'aniu%
\ \protect \BOthers {.}}{%
Loto'aniu%
\ \protect \BOthers {.}}{%
{\protect \APACyear {2022}}%
}]{%
Lotoaniu2022}
\APACinsertmetastar {%
Lotoaniu2022}%
\begin{APACrefauthors}%
Loto'aniu, P\BPBI T\BPBI M.%
, Romich, K.%
, Rowland, W.%
, Codrescu, S.%
, Biesecker, D.%
, Johnson, J.%
\BDBL {}Stevens, M.%
\end{APACrefauthors}%
\unskip\
\newblock
\APACrefYearMonthDay{2022}{}{}.
\newblock
{\BBOQ}\APACrefatitle {Validation of the DSCOVR Spacecraft Mission Space
  Weather Solar Wind Products} {Validation of the dscovr spacecraft mission
  space weather solar wind products}.{\BBCQ}
\newblock
\APACjournalVolNumPages{Space Weather}{20}{10}{e2022SW003085}.
\newblock
\begin{APACrefDOI} \doi{10.1029/2022SW003085} \end{APACrefDOI}
\PrintBackRefs{\CurrentBib}

\bibitem [\protect \citeauthoryear {%
{Manoharan}%
\ \protect \BOthers {.}}{%
{Manoharan}%
\ \protect \BOthers {.}}{%
{\protect \APACyear {2004}}%
}]{%
Manoharan2004}
\APACinsertmetastar {%
Manoharan2004}%
\begin{APACrefauthors}%
{Manoharan}, P\BPBI K.%
, {Gopalswamy}, N.%
, {Yashiro}, S.%
, {Lara}, A.%
, {Michalek}, G.%
\BCBL {}\ \BBA {} {Howard}, R\BPBI A.%
\end{APACrefauthors}%
\unskip\
\newblock
\APACrefYearMonthDay{2004}{}{}.
\newblock
{\BBOQ}\APACrefatitle {{Influence of coronal mass ejection interaction on
  propagation of interplanetary shocks}} {{Influence of coronal mass ejection
  interaction on propagation of interplanetary shocks}}.{\BBCQ}
\newblock
\APACjournalVolNumPages{Journal of Geophysical Research}{109}{A6}{A06109}.
\newblock
\begin{APACrefDOI} \doi{10.1029/2003JA010300} \end{APACrefDOI}
\PrintBackRefs{\CurrentBib}

\bibitem [\protect \citeauthoryear {%
{Mishra}%
\ \BBA {} {Agarwal}%
}{%
{Mishra}%
\ \BBA {} {Agarwal}%
}{%
{\protect \APACyear {2008}}%
}]{%
Mishra2008}
\APACinsertmetastar {%
Mishra2008}%
\begin{APACrefauthors}%
{Mishra}, R\BPBI K.%
\BCBT {}\ \BBA {} {Agarwal}, R.%
\end{APACrefauthors}%
\unskip\
\newblock
\APACrefYearMonthDay{2008}{}{}.
\newblock
{\BBOQ}\APACrefatitle {{Cosmic ray intensity during the passage of coronal mass
  ejections}} {{Cosmic ray intensity during the passage of coronal mass
  ejections}}.{\BBCQ}
\newblock
\APACjournalVolNumPages{Brazilian Journal of Physics}{38}{4}{569-572}.
\newblock
\begin{APACrefDOI} \doi{10.1590/S0103-97332008000500007} \end{APACrefDOI}
\PrintBackRefs{\CurrentBib}

\bibitem [\protect \citeauthoryear {%
{Munakata}%
\ \protect \BOthers {.}}{%
{Munakata}%
\ \protect \BOthers {.}}{%
{\protect \APACyear {2005}}%
}]{%
Munakata2005}
\APACinsertmetastar {%
Munakata2005}%
\begin{APACrefauthors}%
{Munakata}, K.%
, {Kuwabara}, T.%
, {Bieber}, J\BPBI W.%
, {Evenson}, P.%
, {Pyle}, R.%
, {Yasue}, S.%
\BDBL {}{Schuch}, N\BPBI J.%
\end{APACrefauthors}%
\unskip\
\newblock
\APACrefYearMonthDay{2005}{}{}.
\newblock
{\BBOQ}\APACrefatitle {{CME-geometry and cosmic-ray anisotropy observed by a
  prototype muon detector network}} {{CME-geometry and cosmic-ray anisotropy
  observed by a prototype muon detector network}}.{\BBCQ}
\newblock
\APACjournalVolNumPages{Advances in Space Research}{36}{12}{2357-2362}.
\newblock
\begin{APACrefDOI} \doi{10.1016/j.asr.2003.05.064} \end{APACrefDOI}
\PrintBackRefs{\CurrentBib}

\bibitem [\protect \citeauthoryear {%
{Pedersen}%
, {Juusola}%
, {Vanhamäki}%
, {Aikio}%
\BCBL {}\ \BBA {} {Viljanen}%
}{%
{Pedersen}%
\ \protect \BOthers {.}}{%
{\protect \APACyear {2024}}%
}]{%
Pedersen2024}
\APACinsertmetastar {%
Pedersen2024}%
\begin{APACrefauthors}%
{Pedersen}, M\BPBI N.%
, {Juusola}, L.%
, {Vanhamäki}, H.%
, {Aikio}, A\BPBI T.%
\BCBL {}\ \BBA {} {Viljanen}, A.%
\end{APACrefauthors}%
\unskip\
\newblock
\APACrefYearMonthDay{2024}{}{}.
\newblock
{\BBOQ}\APACrefatitle {{Rapid geomagnetic variations during high-speed stream,
  sheath and magnetic cloud-driven geomagnetic storms from 1996 to 2023}}
  {{Rapid geomagnetic variations during high-speed stream, sheath and magnetic
  cloud-driven geomagnetic storms from 1996 to 2023}}.{\BBCQ}
\newblock
\APACjournalVolNumPages{Journal of Geophysical Research: Space
  Physics}{129}{}{e2024JA032656}.
\newblock
\begin{APACrefDOI} \doi{10.1029/2024JA032656} \end{APACrefDOI}
\PrintBackRefs{\CurrentBib}

\bibitem [\protect \citeauthoryear {%
{Richardson}%
}{%
{Richardson}%
}{%
{\protect \APACyear {2018}}%
}]{%
Richardson2018}
\APACinsertmetastar {%
Richardson2018}%
\begin{APACrefauthors}%
{Richardson}, I\BPBI G.%
\end{APACrefauthors}%
\unskip\
\newblock
\APACrefYearMonthDay{2018}{}{}.
\newblock
{\BBOQ}\APACrefatitle {{Solar wind stream interaction regions throughout the
  heliosphere}} {{Solar wind stream interaction regions throughout the
  heliosphere}}.{\BBCQ}
\newblock
\APACjournalVolNumPages{Living Reviews in Solar Physics}{15}{1}{1}.
\newblock
\begin{APACrefDOI} \doi{10.1007/s41116-017-0011-z} \end{APACrefDOI}
\PrintBackRefs{\CurrentBib}

\bibitem [\protect \citeauthoryear {%
{Richardson}%
\ \BBA {} {Cane}%
}{%
{Richardson}%
\ \BBA {} {Cane}%
}{%
{\protect \APACyear {2012}}%
}]{%
Richardson2012}
\APACinsertmetastar {%
Richardson2012}%
\begin{APACrefauthors}%
{Richardson}, I\BPBI G.%
\BCBT {}\ \BBA {} {Cane}, H\BPBI V.%
\end{APACrefauthors}%
\unskip\
\newblock
\APACrefYearMonthDay{2012}{}{}.
\newblock
{\BBOQ}\APACrefatitle {{Solar wind drivers of geomagnetic storms during more
  than four solar cycles}} {{Solar wind drivers of geomagnetic storms during
  more than four solar cycles}}.{\BBCQ}
\newblock
\APACjournalVolNumPages{Journal of Space Weather and Space
  Climate}{2}{A01}{A01}.
\newblock
\begin{APACrefDOI} \doi{10.1051/swsc/2012001} \end{APACrefDOI}
\PrintBackRefs{\CurrentBib}

\bibitem [\protect \citeauthoryear {%
{Schlickeiser}%
, {Oppotsch}%
, {Zhang}%
\BCBL {}\ \BBA {} {Pogorelov}%
}{%
{Schlickeiser}%
\ \protect \BOthers {.}}{%
{\protect \APACyear {2019}}%
}]{%
Schlickeiser2019}
\APACinsertmetastar {%
Schlickeiser2019}%
\begin{APACrefauthors}%
{Schlickeiser}, R.%
, {Oppotsch}, J.%
, {Zhang}, M.%
\BCBL {}\ \BBA {} {Pogorelov}, N\BPBI V.%
\end{APACrefauthors}%
\unskip\
\newblock
\APACrefYearMonthDay{2019}{}{}.
\newblock
{\BBOQ}\APACrefatitle {{On the anisotropy of galactic cosmic rays}} {{On the
  anisotropy of galactic cosmic rays}}.{\BBCQ}
\newblock
\APACjournalVolNumPages{The Astrophysical Journal}{879}{1}{29}.
\newblock
\begin{APACrefDOI} \doi{10.3847/1538-4357/ab24c1} \end{APACrefDOI}
\PrintBackRefs{\CurrentBib}

\bibitem [\protect \citeauthoryear {%
Schrijver%
\ \BBA {} Siscoe%
}{%
Schrijver%
\ \BBA {} Siscoe%
}{%
{\protect \APACyear {2010}}%
}]{%
Schrijver_Siscoe_2010}
\APACinsertmetastar {%
Schrijver_Siscoe_2010}%
\begin{APACrefauthors}%
Schrijver, C\BPBI J.%
\BCBT {}\ \BBA {} Siscoe, G\BPBI L.%
\end{APACrefauthors}%
\unskip\
\newblock
\APACrefYear{2010}.
\newblock
\APACrefbtitle {Heliophysics: Space Storms and Radiation: Causes and Effects}
  {Heliophysics: Space storms and radiation: Causes and effects}\ (C\BPBI
  J.~Schrijver\ \BBA {} G\BPBI L.~Siscoe, \BEDS{}).
\newblock
\APACaddressPublisher{}{Cambridge University Press}.
\newblock
\begin{APACrefDOI} \doi{10.1017/CBO9781139194532} \end{APACrefDOI}
\PrintBackRefs{\CurrentBib}

\bibitem [\protect \citeauthoryear {%
{Schwenn}%
\ \protect \BOthers {.}}{%
{Schwenn}%
\ \protect \BOthers {.}}{%
{\protect \APACyear {2006}}%
}]{%
Schwenn2006}
\APACinsertmetastar {%
Schwenn2006}%
\begin{APACrefauthors}%
{Schwenn}, R.%
, {Raymond}, J\BPBI C.%
, {Alexander}, D.%
, {Ciaravella}, A.%
, {Gopalswamy}, N.%
, {Howard}, R.%
\BDBL {}{Zhang}, J.%
\end{APACrefauthors}%
\unskip\
\newblock
\APACrefYearMonthDay{2006}{}{}.
\newblock
{\BBOQ}\APACrefatitle {{Coronal observations of CMEs}} {{Coronal observations
  of CMEs}}.{\BBCQ}
\newblock
\APACjournalVolNumPages{Space Science Reviews}{123}{1-4}{127-176}.
\newblock
\begin{APACrefDOI} \doi{10.1007/s11214-006-9016-y} \end{APACrefDOI}
\PrintBackRefs{\CurrentBib}

\bibitem [\protect \citeauthoryear {%
{Stone}%
\ \protect \BOthers {.}}{%
{Stone}%
\ \protect \BOthers {.}}{%
{\protect \APACyear {1998}}%
}]{%
Stone1998}
\APACinsertmetastar {%
Stone1998}%
\begin{APACrefauthors}%
{Stone}, E\BPBI C.%
, {Frandsen}, A\BPBI M.%
, {Mewaldt}, R\BPBI A.%
, {Christian}, E\BPBI R.%
, {Margolies}, D.%
, {Ormes}, J\BPBI F.%
\BCBL {}\ \BBA {} {Snow}, F.%
\end{APACrefauthors}%
\unskip\
\newblock
\APACrefYearMonthDay{1998}{}{}.
\newblock
{\BBOQ}\APACrefatitle {{The Advanced Composition Explorer}} {{The Advanced
  Composition Explorer}}.{\BBCQ}
\newblock
\APACjournalVolNumPages{Space Science Reviews}{86}{}{1-22}.
\newblock
\begin{APACrefDOI} \doi{10.1023/A:1005082526237} \end{APACrefDOI}
\PrintBackRefs{\CurrentBib}

\bibitem [\protect \citeauthoryear {%
{Temerin}%
\ \BBA {} {Li}%
}{%
{Temerin}%
\ \BBA {} {Li}%
}{%
{\protect \APACyear {2002}}%
}]{%
Temerin2002}
\APACinsertmetastar {%
Temerin2002}%
\begin{APACrefauthors}%
{Temerin}, M.%
\BCBT {}\ \BBA {} {Li}, X.%
\end{APACrefauthors}%
\unskip\
\newblock
\APACrefYearMonthDay{2002}{}{}.
\newblock
{\BBOQ}\APACrefatitle {{A new model for the prediction of Dst on the basis of
  the solar wind}} {{A new model for the prediction of Dst on the basis of the
  solar wind}}.{\BBCQ}
\newblock
\APACjournalVolNumPages{Journal of Geophysical Research: Space
  Physics}{107}{A12}{1472}.
\newblock
\begin{APACrefDOI} \doi{10.1029/2001JA007532} \end{APACrefDOI}
\PrintBackRefs{\CurrentBib}

\bibitem [\protect \citeauthoryear {%
Tsurutani%
\ \BBA {} Gonzalez%
}{%
Tsurutani%
\ \BBA {} Gonzalez%
}{%
{\protect \APACyear {1997}}%
}]{%
tsurutani1997interplanetary}
\APACinsertmetastar {%
tsurutani1997interplanetary}%
\begin{APACrefauthors}%
Tsurutani, B\BPBI T.%
\BCBT {}\ \BBA {} Gonzalez, W\BPBI D.%
\end{APACrefauthors}%
\unskip\
\newblock
\APACrefYearMonthDay{1997}{}{}.
\newblock
{\BBOQ}\APACrefatitle {The interplanetary causes of geomagnetic storms: A
  review} {The interplanetary causes of geomagnetic storms: A review}.{\BBCQ}
\newblock
\APACjournalVolNumPages{Geophysical Monograph}{98}{}{77--90}.
\newblock
\begin{APACrefDOI} \doi{10.1029/GM098p0077} \end{APACrefDOI}
\PrintBackRefs{\CurrentBib}

\bibitem [\protect \citeauthoryear {%
{Tsurutani}%
\ \protect \BOthers {.}}{%
{Tsurutani}%
\ \protect \BOthers {.}}{%
{\protect \APACyear {1995}}%
}]{%
Tsurutani1995}
\APACinsertmetastar {%
Tsurutani1995}%
\begin{APACrefauthors}%
{Tsurutani}, B\BPBI T.%
, {Gonzalez}, W\BPBI D.%
, {Gonzalez}, A\BPBI L\BPBI C.%
, {Tang}, F.%
, {Arballo}, J\BPBI K.%
\BCBL {}\ \BBA {} {Okada}, M.%
\end{APACrefauthors}%
\unskip\
\newblock
\APACrefYearMonthDay{1995}{}{}.
\newblock
{\BBOQ}\APACrefatitle {{Interplanetary origin of geomagnetic activity in the
  declining phase of the solar cycle}} {{Interplanetary origin of geomagnetic
  activity in the declining phase of the solar cycle}}.{\BBCQ}
\newblock
\APACjournalVolNumPages{Journal of Geophysical
  Research}{100}{A11}{21717-21733}.
\newblock
\begin{APACrefDOI} \doi{10.1029/95JA01476} \end{APACrefDOI}
\PrintBackRefs{\CurrentBib}

\bibitem [\protect \citeauthoryear {%
Uga%
\ \BBA {} Adhikari%
}{%
Uga%
\ \BBA {} Adhikari%
}{%
{\protect \APACyear {2023}}%
}]{%
Uga2023CRI}
\APACinsertmetastar {%
Uga2023CRI}%
\begin{APACrefauthors}%
Uga, C\BPBI I.%
\BCBT {}\ \BBA {} Adhikari, B.%
\end{APACrefauthors}%
\unskip\
\newblock
\APACrefYearMonthDay{2023}{}{}.
\newblock
{\BBOQ}\APACrefatitle {Study of Cosmic Ray Intensity (CRI) along with Solar
  Wind Parameters and Geomagnetic Indices from Different Stations} {Study of
  cosmic ray intensity (cri) along with solar wind parameters and geomagnetic
  indices from different stations}.{\BBCQ}
\newblock
\APACjournalVolNumPages{Cosmic Research}{61}{5}{364--379}.
\newblock
\begin{APACrefDOI} \doi{10.1134/s0010952523600026} \end{APACrefDOI}
\PrintBackRefs{\CurrentBib}

\bibitem [\protect \citeauthoryear {%
{Uga}%
, {Goshu}%
\BCBL {}\ \BBA {} {Rikitu}%
}{%
{Uga}%
\ \protect \BOthers {.}}{%
{\protect \APACyear {2025}}%
}]{%
Uga2025}
\APACinsertmetastar {%
Uga2025}%
\begin{APACrefauthors}%
{Uga}, C\BPBI I.%
, {Goshu}, C\BPBI Y.%
\BCBL {}\ \BBA {} {Rikitu}, K\BPBI S.%
\end{APACrefauthors}%
\unskip\
\newblock
\APACrefYearMonthDay{2025}{}{}.
\newblock
{\BBOQ}\APACrefatitle {{Solar Wind Energy Coupling and Cosmic Ray Intensity: A
  Study of Key Solar Parameters}} {{Solar Wind Energy Coupling and Cosmic Ray
  Intensity: A Study of Key Solar Parameters}}.{\BBCQ}
\newblock
\APACjournalVolNumPages{Radio Science}{60}{}{e2025RS008233}.
\newblock
\begin{APACrefDOI} \doi{10.1029/2025RS008233} \end{APACrefDOI}
\PrintBackRefs{\CurrentBib}

\bibitem [\protect \citeauthoryear {%
Ugwoke%
\ \protect \BOthers {.}}{%
Ugwoke%
\ \protect \BOthers {.}}{%
{\protect \APACyear {2023}}%
}]{%
Ugwoke2023}
\APACinsertmetastar {%
Ugwoke2023}%
\begin{APACrefauthors}%
Ugwoke, R\BPBI E.%
, Ubachukwu, A\BPBI A.%
, Urama, J\BPBI O.%
, Okike, O.%
, Alhassan, J\BPBI A.%
\BCBL {}\ \BBA {} Chukwude, A\BPBI E.%
\end{APACrefauthors}%
\unskip\
\newblock
\APACrefYearMonthDay{2023}{mar}{}.
\newblock
{\BBOQ}\APACrefatitle {Principal Component Analysis of Ground Level Enhancement
  of Cosmic Ray Events} {Principal component analysis of ground level
  enhancement of cosmic ray events}.{\BBCQ}
\newblock
\APACjournalVolNumPages{Research in Astronomy and Astrophysics}{23}{3}{035017}.
\newblock
\begin{APACrefDOI} \doi{10.1088/1674-4527/acba65} \end{APACrefDOI}
\PrintBackRefs{\CurrentBib}

\bibitem [\protect \citeauthoryear {%
{Vr{\v{s}}nak}%
, {Temmer}%
\BCBL {}\ \BBA {} {Veronig}%
}{%
{Vr{\v{s}}nak}%
\ \protect \BOthers {.}}{%
{\protect \APACyear {2007}}%
}]{%
Vrsnak2007}
\APACinsertmetastar {%
Vrsnak2007}%
\begin{APACrefauthors}%
{Vr{\v{s}}nak}, B.%
, {Temmer}, M.%
\BCBL {}\ \BBA {} {Veronig}, A\BPBI M.%
\end{APACrefauthors}%
\unskip\
\newblock
\APACrefYearMonthDay{2007}{}{}.
\newblock
{\BBOQ}\APACrefatitle {{Coronal holes and solar wind high-speed streams: I.
  Forecasting the solar wind parameters}} {{Coronal holes and solar wind
  high-speed streams: I. Forecasting the solar wind parameters}}.{\BBCQ}
\newblock
\APACjournalVolNumPages{Solar Physics}{240}{2}{315-330}.
\newblock
\begin{APACrefDOI} \doi{10.1007/s11207-007-0285-8} \end{APACrefDOI}
\PrintBackRefs{\CurrentBib}

\bibitem [\protect \citeauthoryear {%
{Vr{\v{s}}nak}%
\ \protect \BOthers {.}}{%
{Vr{\v{s}}nak}%
\ \protect \BOthers {.}}{%
{\protect \APACyear {2013}}%
}]{%
Vrsnak2013}
\APACinsertmetastar {%
Vrsnak2013}%
\begin{APACrefauthors}%
{Vr{\v{s}}nak}, B.%
, {{\v{Z}}ic}, T.%
, {Vrbanec}, D.%
, {Temmer}, M.%
, {Rollett}, T.%
, {M{\"o}stl}, C.%
\BDBL {}{Shanmugaraju}, A.%
\end{APACrefauthors}%
\unskip\
\newblock
\APACrefYearMonthDay{2013}{}{}.
\newblock
{\BBOQ}\APACrefatitle {{Propagation of interplanetary coronal mass ejections:
  The drag-based model}} {{Propagation of interplanetary coronal mass
  ejections: The drag-based model}}.{\BBCQ}
\newblock
\APACjournalVolNumPages{Solar Physics}{285}{1-2}{295-315}.
\newblock
\begin{APACrefDOI} \doi{10.1007/s11207-012-0035-4} \end{APACrefDOI}
\PrintBackRefs{\CurrentBib}

\bibitem [\protect \citeauthoryear {%
{Zhang}%
\ \protect \BOthers {.}}{%
{Zhang}%
\ \protect \BOthers {.}}{%
{\protect \APACyear {2007}}%
}]{%
Zhang2007}
\APACinsertmetastar {%
Zhang2007}%
\begin{APACrefauthors}%
{Zhang}, J.%
, {Richardson}, I\BPBI G.%
, {Webb}, D\BPBI F.%
, {Gopalswamy}, N.%
, {Huttunen}, E.%
, {Kasper}, J\BPBI C.%
\BDBL {}{Zhukov}, A\BPBI N.%
\end{APACrefauthors}%
\unskip\
\newblock
\APACrefYearMonthDay{2007}{}{}.
\newblock
{\BBOQ}\APACrefatitle {{Solar and interplanetary sources of major geomagnetic
  storms (Dst $\leq$ -100 nT) during 1996--2005}} {{Solar and interplanetary
  sources of major geomagnetic storms (Dst $\leq$ -100 nT) during
  1996--2005}}.{\BBCQ}
\newblock
\APACjournalVolNumPages{Journal of Geophysical Research}{112}{A10}{A10102}.
\newblock
\begin{APACrefDOI} \doi{10.1029/2007JA012321} \end{APACrefDOI}
\PrintBackRefs{\CurrentBib}

\bibitem [\protect \citeauthoryear {%
{Zhou}%
, {Zhou}%
, {Cao}%
\BCBL {}\ \protect \BOthers {.}}{%
{Zhou}%
, {Zhou}%
, {Cao}%
\BCBL {}\ \protect \BOthers {.}}{%
{\protect \APACyear {2026}}%
}]{%
Zhou2026JGR}
\APACinsertmetastar {%
Zhou2026JGR}%
\begin{APACrefauthors}%
{Zhou}, W.%
, {Zhou}, W.%
, {Cao}, J.%
, {Ge}, Z\BHBI Y.%
, {E}, P.%
\BCBL {}\ \BBA {} {Zhao}, B.%
\end{APACrefauthors}%
\unskip\
\newblock
\APACrefYearMonthDay{2026}{}{}.
\newblock
{\BBOQ}\APACrefatitle {{Evaluating the role of AIA 193 {\AA} solar images
  beyond historical Dst and near-Earth solar wind and interplanetary magnetic
  field information for 1--6 hour Dst forecasting}} {{Evaluating the role of
  AIA 193 {\AA} solar images beyond historical Dst and near-Earth solar wind
  and interplanetary magnetic field information for 1--6 hour Dst
  forecasting}}.{\BBCQ}
\newblock
\APACjournalVolNumPages{Journal of Geophysical Research: Space
  Physics}{131}{}{e2026JA035757}.
\newblock
\begin{APACrefDOI} \doi{10.1029/2026JA035757} \end{APACrefDOI}
\PrintBackRefs{\CurrentBib}

\bibitem [\protect \citeauthoryear {%
{Zhou}%
, {Zhou}%
, {Yang}%
, {Cao}%
\BCBL {}\ \BBA {} {Ge}%
}{%
{Zhou}%
, {Zhou}%
, {Yang}%
, {Cao}%
\BCBL {}\ \BBA {} {Ge}%
}{%
{\protect \APACyear {2026}}%
}]{%
Zhou2026RAA}
\APACinsertmetastar {%
Zhou2026RAA}%
\begin{APACrefauthors}%
{Zhou}, W.%
, {Zhou}, W.%
, {Yang}, Y.%
, {Cao}, J.%
\BCBL {}\ \BBA {} {Ge}, Z.%
\end{APACrefauthors}%
\unskip\
\newblock
\APACrefYearMonthDay{2026}{}{}.
\newblock
{\BBOQ}\APACrefatitle {{Reducing forecast lag in Dst prediction via an Informer
  model with a lag-aware composite loss}} {{Reducing forecast lag in Dst
  prediction via an Informer model with a lag-aware composite loss}}.{\BBCQ}
\newblock
\APACjournalVolNumPages{Research in Astronomy and Astrophysics}{}{}{}.
\newblock
\begin{APACrefDOI} \doi{10.1088/1674-4527/ae903f} \end{APACrefDOI}
\PrintBackRefs{\CurrentBib}

\bibitem [\protect \citeauthoryear {%
{Zhou}%
, {Zhou}%
, {Yang}%
, {Cao}%
, {Ge}%
, {Xu}%
\BCBL {}\ \protect \BOthers {.}}{%
{Zhou}%
, {Zhou}%
, {Yang}%
, {Cao}%
, {Ge}%
, {Xu}%
\BCBL {}\ \protect \BOthers {.}}{%
{\protect \APACyear {2026}}%
}]{%
Zhou2026ApJ}
\APACinsertmetastar {%
Zhou2026ApJ}%
\begin{APACrefauthors}%
{Zhou}, W.%
, {Zhou}, W.%
, {Yang}, Y.%
, {Cao}, J.%
, {Ge}, Z\BHBI Y.%
, {Xu}, T.%
\BDBL {}{Dong}, Y.%
\end{APACrefauthors}%
\unskip\
\newblock
\APACrefYearMonthDay{2026}{}{}.
\newblock
{\BBOQ}\APACrefatitle {{Coronal mass ejection transit time prediction based on
  spatiotemporal modeling using 3D convolutional neural networks}} {{Coronal
  mass ejection transit time prediction based on spatiotemporal modeling using
  3D convolutional neural networks}}.{\BBCQ}
\newblock
\APACjournalVolNumPages{The Astrophysical Journal}{999}{}{23}.
\newblock
\begin{APACrefDOI} \doi{10.3847/1538-4357/ae3c79} \end{APACrefDOI}
\PrintBackRefs{\CurrentBib}

\bibitem [\protect \citeauthoryear {%
{Zhu}%
, {Xue}%
, {Cheng}%
\BCBL {}\ \BBA {} {Cang}%
}{%
{Zhu}%
\ \protect \BOthers {.}}{%
{\protect \APACyear {2015}}%
}]{%
Zhu2015}
\APACinsertmetastar {%
Zhu2015}%
\begin{APACrefauthors}%
{Zhu}, X\BPBI L.%
, {Xue}, B\BPBI S.%
, {Cheng}, G\BPBI S.%
\BCBL {}\ \BBA {} {Cang}, Z\BPBI Y.%
\end{APACrefauthors}%
\unskip\
\newblock
\APACrefYearMonthDay{2015}{}{}.
\newblock
{\BBOQ}\APACrefatitle {{Application of wavelet analysis of cosmic ray in
  prediction of great geomagnetic storms}} {{Application of wavelet analysis of
  cosmic ray in prediction of great geomagnetic storms}}.{\BBCQ}
\newblock
\APACjournalVolNumPages{Chinese Journal of Geophysics}{58}{7}{2242-2249}.
\newblock
\begin{APACrefDOI} \doi{10.6038/cjg20150703} \end{APACrefDOI}
\PrintBackRefs{\CurrentBib}

\end{thebibliography}

\end{document}